\documentclass[aps,prd,twocolumn,preprintnumbers,
superscriptaddress,
showpacs,floatfix]{revtex4}

\usepackage{bm}
\usepackage{epsfig}
\usepackage{graphics}

\newcommand{\eeq}{\end{equation}}
\newcommand{\beq}{\begin{equation}}
\newcommand{\ba}{\begin{array}}
\newcommand{\ea}{\end{array}}
\newcommand{\bea}{\begin{eqnarray}}
\newcommand{\eea}{\end{eqnarray}}
\newcommand{\vev}[1]{\langle #1\rangle}
\newcommand{\vp}{\varphi}
\newcommand{\eps}{\epsilon}

\newcommand{\al}{\alpha}

\begin{document}

\preprint{UT-STPD-2/05}

\title{Non-thermal leptogenesis via direct
inflaton decay without ${\rm SU}(2)_L$
triplets}

\author{Thomas Dent}
\email{tdent@gen.auth.gr}
\author{George Lazarides}
\email{lazaride@eng.auth.gr}
\affiliation{Physics Division, School of
Technology, Aristotle University of
Thessaloniki, Thessaloniki 54124, Greece}
\author{Roberto Ruiz de Austri}
\email{rruiz@delta.ft.uam.es}
\affiliation{Departamento de F\'{\i}sica
Te\'{o}rica C-XI and Instituto de F\'{\i}sica
Te\'{o}rica C-XVI, Universidad Aut\'{o}noma de
Madrid, Cantoblanco, Madrid 28049, Spain}

\date{\today}

\begin{abstract}
\noindent
We present a non-thermal leptogenesis scenario
following standard supersymmetric hybrid
inflation, in the case where light neutrinos
acquire mass via the usual seesaw mechanism and
inflaton decay to heavy right-handed neutrino
superfields is kinematically disallowed, or the
right-handed neutrinos which can be decay
products of the inflaton are unable to generate
sufficient baryon asymmetry via their subsequent
decay. The primordial lepton
asymmetry is generated through the decay of the
inflaton into light particles by the
interference of one-loop diagrams with exchange
of different right-handed neutrinos. The
mechanism requires superpotential couplings
explicitly violating a ${\rm U}(1)$ R-symmetry
and R-parity. We take into account the
constraints from neutrino masses and mixing and
the preservation of the primordial asymmetry.
We consider two models, one without and one
with ${\rm SU}(2)_R$ gauge symmetry. We show
that the former is viable, whereas the latter
is ruled out. Although the broken R-parity need
not have currently observable low-energy
signatures, some R-parity-violating slepton
decays may be detectable in the future
colliders.
\end{abstract}

\maketitle

\section{Introduction}
\label{sec:intro}

\par
One of the most promising scenarios for
generating the observed baryon asymmetry of the
universe (BAU) is certainly the leptogenesis
scenario \cite{lepto,tripletdecay}. It applies
in all the cases where the light neutrinos
($\nu$) acquire their mass by coupling to heavy
standard model (SM) singlet fermions $\nu^c$,
the right-handed neutrinos (RHNs) (this is
known as the seesaw mechanism \cite{seesaw}),
or ${\rm SU}(2)_L$ triplet Higgs scalars
\cite{triplet}. These heavy particles can decay
out of thermal equilibrium generating a
primordial lepton asymmetry, which is
subsequently converted in part into the
observed baryon asymmetry by non-perturbative
sphaleron effects at the electroweak phase
transition.

\par
In the original realization \cite{lepto} of
this scenario, the heavy particles were assumed
to be thermally produced in the early universe.
However, there is a tension between correct
neutrino masses and this thermal leptogenesis
scenario in supersymmetric (SUSY) models
because of the gravitino problem
\cite{khlopov,gravitino}. Assuming that the
gravitino mass is of the order of $1~{\rm TeV}$
and employing generic sparticle spectra, one
finds that the reheat temperature $T_{\rm reh}$
should not exceed about $10^{9}~{\rm GeV}$
since otherwise an unacceptably large number
density of gravitinos is thermally produced at
reheating. These gravitinos later decay
presumably into photons and photinos
interfering with the successful predictions of
standard big bang nucleosynthesis. On the other
hand, adequate thermal production of RHNs or
${\rm SU}(2)_L$ triplets, whose decay creates
the primordial lepton asymmetry, requires that
the mass of these particles does not exceed
$T_{\rm reh}$. This leads to unacceptably large
light neutrino masses. This problem can be
alleviated \cite{pilaftsis,deg} if we allow
some degree of degeneracy between the relevant
RHNs, which enhances the generated lepton
asymmetry, and perhaps also if the branching
ratio of the gravitino decay into photons and
photinos is less than unity, which somewhat
relaxes \cite{gravitino} the gravitino
constraint on $T_{\rm reh}$.

\par
A much more natural solution of the tension
between the gravitino bound on $T_{\rm reh}$
and the masses of light neutrinos is provided
by non-thermal leptogenesis \cite{inflepto} at
reheating. In existing realizations
\cite{nonthtripletdec} of this scenario,
though, where the inflaton decays into RHN or
${\rm SU}(2)_L$ triplet superfields, this still
puts a restriction on the masses of these
particles: the decay products of the inflaton
must be lighter than half its mass
$m_{\rm inf}$. The primordial lepton asymmetry
is generated in the subsequent decay of the RHN
or ${\rm SU}(2)_L$ triplet superfields.

\par
In a recent paper \cite{previous}, we
considered the consequences of allowing all the
RHN and ${\rm SU}(2)_L$ triplet superfields
to be heavier than $m_{\rm inf}/2$ (see also
Ref.~\cite{allahv}). Primordial leptogenesis
could then take place only through the direct
decay of the inflaton into light
particles (see also Ref.~\cite{raidal}). We
took a simple SUSY grand unified theory (GUT)
model which is based on the gauge group
$G_{B-L}=G_{\rm SM}\times{\rm U}(1)_{B-L}$
($G_{\rm SM}$ is the SM gauge group, and $B$
and $L$ the baryon and lepton number
respectively) and naturally incorporates the
standard SUSY realization \cite{cllsw,dss} of
hybrid inflation \cite{hybrid}. The flatness of
the inflationary trajectory at tree level was
guaranteed by a ${\rm U}(1)$ R-symmetry,
whereas radiative corrections provided
\cite{dss} a logarithmic slope along
this path, needed for driving the inflaton
towards the SUSY vacua. The R-symmetry also
guaranteed the conservation of baryon number
to all orders in perturbation theory.
Therefore, baryon number was only violated by
the non-perturbative electroweak sphaleron
effects.

\par
The model incorporated the solution of the
$\mu$ problem of the minimal supersymmetric
standard model (MSSM) proposed in
Ref.~\cite{lr}. Although the global
R-symmetry forbade the appearance of a $\mu$
term in the superpotential, it did allow
the existence of the trilinear term $Sh_1h_2$,
where $S$ is the gauge singlet inflaton of
standard SUSY hybrid inflation and $h_1$,
$h_2$ are the electroweak Higgs superfields.
After
the GUT gauge symmetry breaking, the soft
SUSY-breaking terms, which generally violated
the R-symmetry, gave rise to a suppressed
linear term in $S$ and, thus, this field
acquired a vacuum expectation value (VEV) of
the order of the electroweak scale divided by a
small coupling constant. The above trilinear
coupling could then yield a $\mu$ term of the
right magnitude.

\par
The same coupling also gave rise to tree-level
couplings of the inflaton, which consisted of
two complex scalar fields, to the electroweak
Higgs bosons and Higgsinos. After the
termination of inflation, the inflaton performs
damped oscillations about the SUSY vacuum and
eventually decays predominantly into
electroweak Higgs superfields via these
tree-level couplings, thereby reheating the
universe. The model contained both heavy RHN
and ${\rm SU}(2)_L$ triplet superfields which
both contributed to light neutrino masses. A
primordial lepton asymmetry could be generated
at reheating via the subdominant decay of the
inflaton into lepton and Higgs superfields
through the interference between one-loop
diagrams with RHN and ${\rm SU}(2)_L$ triplet
exchange respectively. The simultaneous
presence of both RHNs and ${\rm SU}(2)_L$
triplets was essential for this particular
leptogenesis mechanism to work. Note that, in
the model of Ref.~\cite{previous}, the
generation of a non-zero lepton asymmetry did
not rely on the existence of more than one
fermion family. However, more than one family
was required if the lepton asymmetry was to be
preserved down to the electroweak phase
transition. Finally, it should be emphasized
that the generation of a non-zero lepton
asymmetry required the inclusion of some
couplings in the superpotential that explicitly
violate the ${\rm U}(1)$ R-symmetry.

\par
In this paper, we investigate the consequences
of doing without ${\rm SU}(2)_L$ triplet
superfields and instead generating the
primordial lepton asymmetry through the
exchange of different RHN superfields. Since
these heavy fields can only appear in
intermediate states of the inflaton decay, we
must create the asymmetry directly from this
decay. Indeed, leptogenesis can again occur in
the subdominant decay of the inflaton into
lepton and Higgs superfields, but now through
the interference between one-loop diagrams
with different RHN exchange. The lepton
asymmetry is proportional to a novel
CP-violating invariant product of coupling
constants.

\par
We study this new leptogenesis scenario within
the framework of the model of
Ref.~\cite{previous} with the ${\rm SU}(2)_L$
triplet superfields removed. All the salient
properties of the model are retained except
that now the light neutrino masses are
generated by the standard seesaw mechanism
\cite{seesaw} which involves only RHN
(super)fields and leptogenesis takes place via
the new mechanism mentioned above. In
particular, the implementation of the standard
SUSY hybrid inflationary scenario
\cite{cllsw,dss} and the solution of the $\mu$
problem of Ref.~\cite{lr} remains unaffected.
Also, baryon number is still conserved to all
orders in perturbation theory.

\par
The model again contains an approximate
${\rm U}(1)$ R-symmetry explicitly broken by
some superpotential operators. These operators,
which are necessary to create a non-zero lepton
asymmetry, also violate the $Z_2$ matter parity
subgroup of the R-symmetry which is preserved
by soft SUSY-breaking terms. The matter parity
violation may have important observable
consequences at low energy. Indeed, if the
lightest sparticle (LSP), which is assumed to
be the lightest neutralino, contains a Higgsino
component, we find that it could decay
predominantly into one or two Higgs bosons and
a lepton. These dominant decay channels can
though be easily blocked kinematically if the
LSP is not too heavy and, under certain
conditions, this particle can be made
long-lived (see Sec.~\ref{sec:rsym}). However,
some matter-parity-violating slepton decays
which may be detectable in the future colliders
are typically present (see
Sec.~\ref{sec:numerics}).

\par
We perform a detailed chemical potential
analysis of the evolution of the primordial
lepton asymmetry until the time at which the
baryon-number-violating electroweak sphaleron
effects cease to operate and evaluate the
baryon asymmetry at this moment, which yields
the observed BAU. We find that there are some
lepton-number-violating four-scalar processes
involving Higgs bosons and sleptons which are
in equilibrium in a range of temperatures just
above the SUSY threshold. These processes
(which result from the same dimension-four
operators that cause the above-mentioned
slepton decays) could readily erase the
primordial lepton asymmetry, which would lead
to the absence of any baryon asymmetry in the
present universe. However, it is possible to
choose parameters so that one of the three
lepton family numbers is conserved by all
dimension-four processes. One can further show
that all the dimension-five processes which
violate this number are well out of equilibrium
at all temperatures after reheating. Thus, we
can obtain a non-zero BAU at present.

\par
We find that the value of the BAU from the
Wilkinson microwave anisotropy probe (WMAP)
data \cite{wmap} can be achieved given
constraints from other observables, notably the
reheat temperature and neutrino masses and
mixing, and CP-violating phases of order unity.
Note that, contrary to Ref.~\cite{previous}
where we worked in the two heaviest family
approximation, in the present paper we take
fully into account all three neutrino species.
We find that only for a small range of values
of the lightest neutrino mass $m_1$ can the
scenario be successful, given the requirement
that the GUT breaking scale be perturbative
relative to an underlying string or quantum
gravity scale, which is also restricted to
smaller values than the value used in
Ref.~\cite{previous}. The reheat temperature is
pushed to higher values relative to this
reference. Finally, the values of the neutrino
Yukawa couplings are also restricted. These
results reflect the fact that removing the
${\rm SU}(2)_L$ triplets from the model makes
it more restrictive. The prediction for the
spectral index of density perturbations is
typical of SUSY hybrid inflation models (see
{\it e.g.} Ref.~\cite{lectures}).

\par
Thus, an acceptable value of the baryon
asymmetry can be obtained within a consistent
model of cosmology and particle physics,
without requiring additional fine-tuned
coupling constants, while respecting
experimental constraints on neutrino masses and
mixing. Moreover, although the scenario
requires violation of the R-symmetry, it is not
necessary to introduce superpotential terms
which would lead to currently observable
R-symmetry-violating effects.

\par
We also analyzed the case where ${\rm SU}(2)_R$
gauge symmetry is imposed, forcing the neutrino
Yukawa couplings to be identical to those of
the charged leptons at the GUT scale. This case
is thus considerably more restrictive than the
non-${\rm SU}(2)_R$-symmetric case. One of the
main differences is that it predicts that the
lightest RHN mass is of order $10^7~{\rm GeV}$,
whereas the inflaton mass is greater than
$10^{11}~{\rm GeV}$, thus the inflaton can
decay directly to the lightest RHN. However,
this does not alter our basic picture for
leptogenesis since the corresponding branching
ratio and non-thermally generated lepton
asymmetry are negligible due to the fact that
the lightest RHN mass is very small compared to
the inflaton mass. Moreover, thermal
leptogenesis from the
decay of this particle can be ignored for the
same reason. The presence of the lightest RHN
in the thermal bath after reheating, however,
yields extra restrictions on the parameters of
the model from the requirement that the initial
lepton asymmetry is not erased. In the end,
this case is ruled out because it always yields
an unacceptably small BAU. The reason is that
our leptogenesis scenario requires strong
mixing in the RHN sector but, in this case, the
RHN mass matrix has too small off-diagonal
elements due to the strong hierarchy of Yukawa
coupling constants implied by the
${\rm SU}(2)_R$ symmetry and the extra
restrictions from the preservation of the
primordial asymmetry.

\par
In Sec.~\ref{sec:model}, we introduce the SUSY
GUT models and describe some of their salient
features. In Sec.~\ref{sec:numass}, we discuss
constraints from the light neutrino masses and
mixing, while, in Sec.~\ref{sec:invariants}, we
present the CP-violating invariant products of
coupling constants which enter into the
primordial lepton asymmetry. The calculation of
the primordial lepton asymmetry is sketched in
Sec.~\ref{sec:bau}, and the effects of
R-symmetry (and R-parity) violation are
discussed in Sec.~\ref{sec:rsym}.
Section~\ref{sec:preserve} is devoted to the
conditions for the primordial lepton asymmetry
to be preserved at high temperatures. We
revisit the relation between the initial lepton
asymmetry and the final baryon asymmetry in
Sec.~\ref{sec:chempot}, finding some novel
results for the case when some lepton family
numbers are violated. Our numerical results
appear in Sec.~\ref{sec:numerics}, including a
discussion of why the ${\rm SU}(2)_R$-symmetric
case is ruled out. Finally, conclusions appear
in Sec.~\ref{sec:concl}.

\begin{table}
\caption{${\rm U}(1)$ charges of superfields
\label{tab:charges}}
\begin{ruledtabular}
\begin{tabular}{c|ccccccccccccc}
& $S$ & $\phi$ & $\bar{\phi}$ & $h_1$ &
$h_2$ & $l$ & $\nu^c$ & $e^c$ & $q$
& $u^c$ & $d^c$ \\ \hline
$B-L$ & 0 & 1 & -1 & 0 & 0 & -1 & 1 & 1
& 1/3 & -1/3 & -1/3 \\
$R$ & 2 & 0 & 0 & 0 & 0 & 1 & 1 & 1 & 1
& 1 & 1
\end{tabular}
\end{ruledtabular}
\end{table}

\section{The SUSY GUT models}
\label{sec:model}

\par
We consider two SUSY GUT models which are
based on the gauge groups $G_{B-L}$ and $G_{LR}
={\rm SU}(3)_{c}\times{\rm SU}(2)_{L}\times
{\rm SU}(2)_{R}\times{\rm U}(1)_{B-L}$
respectively. The models also possess an
approximate global R-symmetry ${\rm U}(1)_R$,
which is explicitly broken by some terms in
the superpotential.

\par
We will treat the $G_{B-L}$ case first. In
addition to the usual MSSM chiral superfields
$h_1$, $h_2$ (Higgs ${\rm SU}(2)_L$ doublets),
$l_i$ (${\rm SU}(2)_L$ doublet leptons),
$e^c_i$ (${\rm SU}(2)_L$ singlet charged
leptons), $q_i$ (${\rm SU}(2)_L$ doublet
quarks), and $u^c_i$, $d^c_i$ (${\rm SU}(2)_L$
singlet anti-quarks) with $i=1,2,3$ being the
family index, the model also contains the SM
singlet chiral superfields $\nu^c_i$ (RHNs),
$S$, $\phi$, and $\bar{\phi}$. The charges
under ${\rm U}(1)_{B-L}$ and ${\rm U}(1)_R$
are given in Table~\ref{tab:charges}. The
superpotential is
\bea
W &=&\kappa S(\bar{\phi}\phi-M^2)+
\lambda S(h_1h_2)+y_{eij}(l_ih_1)e^c_j
\nonumber \\
& &+y_{uij}(q_ih_2)u^c_j+y_{dij}(q_ih_1)d^c_j
\nonumber\\
& &+y_{\nu ij}(l_ih_2)\nu^c_j+(M_{\nu^cij}/M^2)
\bar{\phi}^2\nu^c_i\nu^c_j
\nonumber\\
& &+(\lambda_i/M_S)\bar{\phi}\nu^c_i(h_1 h_2)
+\cdots,
\label{W}
\eea
where $M$ is a mass parameter of order
the GUT scale, $M_S$ is the string or quantum
gravity scale $\sim 10^{17}~{\rm GeV}$,
and $(X Y)$ indicates the ${\rm SU}(2)_L$
invariant product $\eps_{ab}X_aY_b$ with
$\eps$ denoting the $2\times 2$ antisymmetric
matrix with $\eps_{12}=1$.
The ellipsis represents terms of order higher
than four and summation over indices is
implied. The only ${\rm U}(1)_R$-violating
couplings which we allow in the superpotential
are the ones in the last explicitly displayed
term in the right hand side (RHS) of
Eq.~(\ref{W}), which are necessary for
leptogenesis. It can be shown that baryon
number ($B$) is automatically conserved to all
orders as a consequence of ${\rm U}(1)_R$. The
argument goes as in Ref.~\cite{nonthtripletdec}
and is not affected by the presence of the
above ${\rm U}(1)_R$-breaking superpotential
couplings. Lepton number ($L$) is then also
conserved in the superpotential as implied by
the presence of ${\rm U}(1)_{B-L}$.

\par
In the ${\rm SU}(2)_R$-symmetric case, the
field content is simplified as follows: the
${\rm SU}(2)_L$ singlets $\nu^c_i$ and $e^c_i$
form the ${\rm SU}(2)_R$ doublets $l^c_i$, the
anti-quark superfields $u^c_i$ and $d^c_i$ form
the ${\rm SU}(2)_R$ doublets $q^c_i$, and the
electroweak Higgs superfields $h_1$ and $h_2$
form the ${\rm SU}(2)_L\times{\rm SU}(2)_R$
bi-doublet $h=(h_2,h_1)$. The symmetry-breaking
singlets $\phi$ and $\bar{\phi}$ are replaced
by ${\rm SU}(2)_R$ doublets $L^c$ and
$\bar{L}^c$ whose neutral (RHN-like) components
$N^c$ and $\bar{N}^c$ obtain VEVs of order the
GUT scale. To simplify notation, we will denote
these components also by $\phi$ and $\bar{\phi}$
respectively. The superfields $L^c$ and
$\bar{L}^c$ have the same charges under
${\rm U}(1)_{B-L}$ and ${\rm U}(1)_R$ as $\phi$
and $\bar{\phi}$ respectively, while the charges
of $l^c_i$, $q^c_i$ and $h$ coincide with the
charges of their components in
Table~\ref{tab:charges}. Below the scale $M$,
the field content is the same as in the previous
case. The superpotential, in this case, is
\bea
W &=&\kappa S(\bar{L}^cL^c-M^2)+\lambda Sh^2+
y_{lij}l_i\eps h l^c_j
\nonumber \\
& &+y_{qij}q_i\eps h q^c_j+(M_{\nu^cij}/M^2)
\bar{L}^cl^c_i\bar{L}^cl^c_j
\nonumber\\
& &+(\lambda_i/M_S)\bar{L}^cl^c_ih^2+\cdots,
\label{WSU2R}
\eea
where $h^2$ denotes the gauge invariant sum
$\frac{1}{2} {\rm Tr}(\eps h^T \eps h)=
(h_1h_2)$ and $L^c$, $l^c_i$, $q^c_i$ are taken
as column 2-vectors, while $\bar{L}^c$, $l_i$,
$q_i$ as row 2-vectors (the superscript $T$
denotes matrix transposition). Again the only
${\rm U}(1)_R$-violating superpotential
couplings are the ones in the last explicitly
displayed term in the RHS of Eq.~(\ref{WSU2R})
and $B$ conservation to all orders is
automatic. Note that, imposing the
${\rm SU}(2)_R$ symmetry on the superpotential
of Eq.~(\ref{W}), we obtain that the Yukawa
coupling constants $y_{eij}$ ($y_{uij}$) and
$y_{\nu ij}$ ($y_{dij}$) become equal
\cite{mixing}. Their common value is the
coupling constant $y_{lij}$ ($y_{qij}$) in
Eq.~(\ref{WSU2R}).

\par
In both cases, the classically flat direction
in field space along which inflation takes
place is as described in Ref.~\cite{lr}: for
$\kappa<\lambda$, it is parametrized by $S$,
$|S|>S_c=M$, with the values of all the other
fields being equal to zero, and has a constant
potential energy density $\kappa^2M^4$ (at
tree level). Here, the dimensionless coupling
constants $\kappa$, $\lambda$ and the mass
parameter $M$ are taken real and positive
by appropriately redefining the phases of the
superfields. There are radiative corrections
\cite{dss} which lift the flatness of this
classically flat direction leading to
slow-roll inflation which, for $\kappa\ll 1$,
terminates practically when $|S|$ reaches the
instability point at $|S|=M$ as one can easily
deduce from the slow-roll $\epsilon$ and $\eta$
criteria \cite{lectures}. The quadrupole
anisotropy of the cosmic microwave background
radiation (CMBR) and the number of e-foldings
\cite{lectures}
\beq
N_Q\simeq\ln\left[1.88\times 10^{11}
\kappa^{\frac{1}{3}}\left(\frac{M}{{\rm GeV}}
\right)^{\frac{2}{3}}\left(\frac{T_{\rm reh}}
{{\rm GeV}}\right)^{\frac{1}{3}}\right]
\label{nq}
\eeq
suffered by our present horizon scale during
inflation are given by the Eqs.~(2)-(4) of
Ref.~\cite{atmo}; in the
non-${\rm SU}(2)_R$-symmetric case, the two
last terms in the RHS of Eq.~(3) must be
divided by two since the ${\rm SU}(2)_R$
doublet chiral superfields $L^c$, $\bar{L}^c$
are replaced by the SM singlets $\phi$,
$\bar\phi$ (see also Ref.~\cite{atmotalk}).

\par
For $\kappa<\lambda$, a GUT-symmetry-breaking
phase transition takes place when the value of
$|S|$ falls below the mass parameter $M$. The
fields evolve towards the realistic SUSY
minimum at $\vev{S}=0$, $\vev{\phi}=
\vev{\bar{\phi}}=M$, $\vev{h_1}=\vev{h_2}=0$,
where the GUT-symmetry-breaking VEVs are taken
real and positive by a $B-L$ rotation. There is
also an unrealistic SUSY minimum which is given
below. With the addition of soft SUSY-breaking
terms, as required in a realistic model, the
position of the vacuum shifts \cite{lr} to
non-zero $\vev{S}\simeq -m_{3/2}/\kappa$, where
$m_{3/2}$ is the mass of the gravitino, and an
effective $\mu$ term with $\mu\simeq -\lambda
m_{3/2}/\kappa\sim m_{3/2}$ is generated from
the superpotential coupling
$\lambda S(h_1h_2)$.

\par
After this phase transition, the (complex)
inflaton degrees of freedom are $S$ and $\theta
\equiv(\delta\phi+\delta\bar{\phi})/\sqrt{2}$,
where $\delta\phi=\phi-M$ and $\delta\bar{\phi}
=\bar{\phi}-M$, with mass $m_{\rm inf}=\sqrt{2}
\kappa M$. These fields oscillate about the
minimum of the potential
and decay to MSSM degrees of freedom reheating
the universe. The predominant decay channels of
$S^*$ and $\theta$ are to fermionic and bosonic
$h_1$, $h_2$ respectively via tree-level
couplings derived from the superpotential terms
$\lambda S(h_1h_2)$ and $\kappa S\bar{\phi}
\phi$ (or $\kappa S\bar{L}^cL^c$). Note that,
if $\kappa>\lambda$, the system would end up in
the unrealistic SUSY minimum at $\phi=\bar\phi=
0$, $|h_1|=|h_2|\simeq(\kappa/\lambda)^{1/2}M$,
which is degenerate with the realistic one (up
to $m_{3/2}^4$) and is separated from it by a
potential barrier of order $m_{3/2}^2M^2$.

\par
The RHNs acquire masses $M_{\nu^c ij}$ after
the spontaneous breaking of ${\rm U}(1)_{B-L}$
by $\vev{\phi}$, $\vev{\bar{\phi}}$. The
coupling constants of the relevant terms in the
third line of the RHS of Eq.~(\ref{W}) (or the
second line of Eq.~(\ref{WSU2R})) can also be
written as $\lambda_{\nu^c ij}/M_S$, making it
clear that the RHN masses are suppressed by a
factor $M/M_S$ relative to $M$. It is possible
to redefine the superfields $\nu^c_i$ (or
$l^c_i$) to obtain effective mass terms $M_i
\nu^c_i\nu^c_i$ which are diagonal in the
flavor space with $M_i$ real and positive.
This basis, which we will call ``RHN basis'',
is most convenient for calculating the BAU (see
Sec.~\ref{sec:bau}). For definiteness, we take
$M_1\leq M_2\leq M_3$.

\par
We will also use a basis in which the lepton
Yukawa couplings and the ${\rm SU}(2)_L$
interactions are diagonal in flavor space,
denoting couplings and fields in this
``lepton family basis'' with a hat. We thus
have $\hat{y}_{eij}=\delta_{ij}\hat{y}_{ei}$,
$\hat{y}_{\nu ij}=\delta_{ij}\hat{y}_i$
(or $\hat{y}_{lij}=\delta_{ij}\hat{y}_{li}$).
The diagonal Yukawa coupling constants
$\hat{y}_{ei}$, $\hat{y}_i$ (or
$\hat{y}_{li}$) are taken real and positive by
appropriate rephasing of the fields. Note that,
in the non-${\rm SU}(2)_R$-symmetric case, it
is an additional assumption that the neutrino
and charged lepton Yukawa couplings can be
diagonalized in the same weak interaction
basis. We will elaborate on this assumption in
Sec.~\ref{sec:preserve}. Since the weak
interactions are in equilibrium at temperatures
above the critical temperature $T_c$ for the
electroweak phase transition, the lepton family
numbers $L_i$ can only be defined in this basis.

\par
Relative to the ``unhatted'' RHN basis, we have
\beq
\hat{l}= l U,\quad \hat{\nu}^c=U^c\nu^c.
\eeq
Here $U$, $U^c$ are $3\times 3$ unitary
matrices and we write ``left-handed'' lepton
superfields, {\it i.e.} ${\rm SU}(2)_L$ doublet
leptons, as row 3-vectors in family space and
``right-handed'' anti-lepton superfields,
{\it i.e.} ${\rm SU}(2)_L$ singlet anti-leptons,
as column 3-vectors. In the
${\rm SU}(2)_R$-symmetric case, we have the
same relation except that the ${\rm SU}(2)_R$
doublet $l^c$ ($\hat{l}^c$) replaces $\nu^c$
($\hat{\nu}^c$). Then by definition we have,
for the neutrino Yukawa couplings,
\bea
l\eps h_2 y_{\nu}\nu^c&=&\hat{l}\eps h_2
U^\dag y_{\nu} U^{c\dag}\hat{\nu}^c,
\nonumber\\
U^\dag y_{\nu} U^{c\dag}&=&\hat{y}_\nu =
{\rm diag}(\hat{y}_1,\hat{y}_2,\hat{y}_3),
\eea
where $y_\nu$ is the $3\times 3$ matrix with
elements $y_{\nu ij}$. In general, the
$\hat{y}_i$ can be any (real positive) numbers.
In the ${\rm SU}(2)_R$-symmetric case, however,
they are determined by the ``asymptotic''
values (at the GUT scale) of the charged lepton
masses as $\hat{y}_{1,2,3}=m_{e,\mu,\tau}\tan
\beta/\vev{h_2}$. The Higgs VEV is $\vev{h_2}
\simeq 174\sin\beta~{\rm GeV}$, which, in the
large $\tan\beta$ limit, yields $\vev{h_2}
\simeq 174~{\rm GeV}$. Exact ${\rm SU}(2)_R$
Yukawa coupling relations in the quark sector
would fix the ratio of the Higgs VEVs, but at
least small deviations are required here to be
consistent with the data. Assuming that the
corrections are small compared to the third
generation Yukawa couplings, $\tan\beta$ should
be about $55$. (For the conditions under which
${\rm SU}(2)_R$ implies large $\tan\beta$, see
Ref.~\cite{largetanb}.) The matrices $U$ and
$U^c$ are, at this stage, determined only up to
a diagonal matrix of arbitrary complex phases
$P={\rm diag}(e^{i\vp_1},e^{i\vp_2},e^{i\vp_3})$
which acts as $U\rightarrow UP$,
$U^c\rightarrow P^{-1}U^c$, corresponding to
opposite phase redefinitions of the lepton weak
doublet and singlet superfields.

\par
As for the ${\rm U}(1)_R$-violating terms,
after the ${\rm U}(1)_{B-L}$ (and
${\rm SU}(2)_R$) breaking, the explicitly
displayed terms in the last line of the RHS of
Eq.~(\ref{W}) or (\ref{WSU2R}) give rise to
effective $B-L$ and matter-parity-violating
operators $\zeta_i\nu^c_i(h_1h_2)$, where
the dimensionless effective coupling constant
$\zeta_i$ is suppressed by one power of
$M/M_S$. If we require that the magnitude
of the dimensionless coupling constants
$\lambda_i$ be less than unity, we obtain a
bound $|\zeta_i|\leq M/M_S$. Note that the
effective coupling constants $\zeta_i$ cannot
in general be made real and positive by
redefining the complex phases of the
superfields. This can be easily shown by
considering the phase redefinition invariant
$\zeta_i^{*2}\mu^2 M_i$ (no summation over
repeated indices) with $\mu$ and $M_i$ already
made real (recall that the $\zeta_i$ are in the
RHN mass eigenstate basis). The coupling
constants $\zeta_i$, thus, remain in general
complex. Note that $\vev{h_1}$, $\vev{h_2}$ can
be taken real because of the reality of
$B\mu\simeq -2\lambda m_{3/2}^2/\kappa$
\cite{lr}. It is, of course, possible to write
down many other R-symmetry-violating operators.
However, they are not necessary for a non-zero
primordial lepton asymmetry to be created.

\par
The Yukawa coupling constants $y_{\nu ij}$ (in
the original RHN mass eigenstate basis) remain
in general complex as one can easily deduce
from the rephasing invariants $y_{\nu ij}
y_{\nu ik}^*\zeta_j^*\zeta_k$ for $j\neq k$ (no
summation over repeated indices), as well as
the standard rephasing invariant $y_{\nu ij}
y_{\nu kj}y_{\nu im}^*y_{\nu km}^*M_j^*M_m$ for
$j\neq m$ (no summation), which has been used
in the original leptogenesis scenario of
Ref.~\cite{lepto}.

\par
The calculation of lepton asymmetry produced
in $S$ and $\theta$ decays is quite
straightforward but somewhat lengthy, and
differs in detail from the usual case where
leptogenesis takes place through the decay
of RHN or ${\rm SU}(2)_L$ triplet
superfields. Since we consider the
interference of two one-loop diagrams, we will
need to find the real parts of loop integrals,
which require renormalization. Before
proceeding to the calculation, we summarize the
constraints derived from the current neutrino
experiments.

\section{Neutrino masses and constraints}
\label{sec:numass}

\par \noindent
The standard seesaw mechanism yields the light
neutrino mass matrix:
\beq
m=-\vev{h_2}^2y_{\nu}M_{\nu^c}^{-1}y_{\nu}^T,
\label{numass}
\eeq
which holds in any basis of neutrino states
with $M_{\nu^c}$ being the $3\times 3$ matrix
with elements $M_{\nu^cij}$. In the RHN mass
eigenstate basis, where $M_{\nu^c}={\rm diag}
(M_1,M_2,M_3)$, it can be rewritten as
\beq
m=-\vev{h_2}^2U\hat{y}_\nu U^cM_{\nu^c}^{-1}
U^{cT}\hat{y}_\nu U^T.
\label{eq:rhnseesaw}
\eeq
Thus, transforming the light neutrino fields to
the ``hatted'' (lepton family) weak interaction
basis, we have
\beq
\hat{m}=U^\dag mU^\ast=-\vev{h_2}^2
\hat{y}_\nu U^cM_{\nu^c}^{-1} U^{cT}
\hat{y}_\nu.
\label{eq:hatseesaw}
\eeq
The light neutrino mass eigenstate basis can be
denoted by $\bar{\nu}$ and is given by
\beq
\bar{\nu}= \hat{\nu}V^\ast,
\eeq
where $\hat{\nu}$, $\bar{\nu}$ are row
3-vectors in family space and $V$ is the
unitary Maki-Nakagawa-Sakata (MNS) matrix which
satisfies
\beq
V^T\hat{m}V=\bar{m}
\equiv {\rm diag}(m_1,m_2,m_3)
\label{eq:mbarmhat}
\eeq
with $m_i$ real and positive. Again $V$ is
determined only up to a diagonal matrix of
complex phases as $V\rightarrow PV$, which
reflects the freedom to redefine the phases of
the superfields in the ``hatted'' basis.
However, the ambiguity is fixed by requiring
$V$ to take the standard form
\beq
V=\left[\ba{ccc} c_3c_2 & s_3c_2 & s_2
e^{-i\delta} \\
-c_1s_3-c_3s_1s_2e^{i\delta} & c_3c_1
-s_3s_1s_2e^{i\delta} & c_2s_1 \\
s_3s_1-c_3c_1s_2e^{i\delta} & -c_3s_1
-c_1s_3s_2e^{i\delta} & c_2c_1 \ea
\right]\cdot\cal{P}
\eeq
where $c_i\equiv\cos\theta_i$, $s_i\equiv\sin
\theta_i$ with $\theta_i$ ($i=1,2,3$) being
the appropriate mixing angles, $\delta$ is
the Dirac phase and ${\cal{P}}={\rm diag}
(e^{-i\al},e^{-i\beta},1)$ contains the
Majorana phases.

\par
In both the ${\rm SU}(2)_R$- and
non-${\rm SU}(2)_R$-symmetric cases, we will
restrict the light neutrino mass-squared
differences and mixing angles to be within the
$2\sigma$ allowed region detailed in
Ref.~\cite{maltoni}. For most numerical
results, however, we will set these observables
to their best-fit values, implying that
$\hat{m}$ is completely determined once the
lightest neutrino mass, the phases $\alpha$ and
$\beta$ and the choice of normal or inverted
hierarchy are fixed. We will take here the
light neutrino mass ordering $m_1\leq m_2\leq
m_3$ and, unless stated otherwise, adopt the
normal hierarchical scheme of neutrino masses,
where the solar and atmospheric neutrino
mass-squared differences are identified with
$\delta m_{21}^2$ and $\delta m_{31}^2$
respectively (here $\delta m_{ij}^2=m_i^2-
m_j^2$). For simplicity, we further take
$\alpha=\beta=0$.

\par
For given values of $m_i$ and $V$, one can
find the light neutrino mass matrix in the
``hatted'' basis by inverting
Eq.~(\ref{eq:mbarmhat}). Then, in the
${\rm SU}(2)_R$-symmetric case, the seesaw
formula in Eq.~(\ref{eq:hatseesaw}) can be
applied with $\hat{y}_i$ determined by the
known values of the charged lepton masses,
renormalization group (RG) evolved to the GUT
scale, and for given $\tan\beta$. Solving the
resulting equations, we can simultaneously
determine the elements of the unitary matrix
$U^c$ and the RHN mass eigenvalues $M_i$. In
the non-${\rm SU}(2)_R$-symmetric case, the
Yukawa coupling constants $\hat{y}_i$ are not
related to the charged lepton masses and,
thus, are treated as free input parameters.

\par
For our mechanism to be the exclusive source
of baryon asymmetry as desired, either all RHN
masses must be greater than $m_{\rm inf}/2$ so
that the inflaton decay to RHN superfields is
kinematically blocked and the thermal
production of RHNs after reheating is
suppressed, or any RHNs that are lighter must
contribute a negligible net lepton number
density via the conventional thermal or
non-thermal leptogenesis mechanism, by which
they decay out of equilibrium. One way to
achieve the latter is to make them adequately
light. We also have to check that thermal
interactions of any lighter RHNs do not wash
out the already created lepton asymmetry.
Furthermore, assuming that the dimensionless
coefficients $\lambda_{\nu^c ij}$ of the last
superpotential terms in the third line of the
RHS of Eq.~(\ref{W}) (or the second line of the
RHS of Eq.~(\ref{WSU2R})) do not exceed unity,
we obtain that the heaviest RHN must not have a
mass greater than $M^2/M_S$. These requirements
place non-trivial constraints on the model
parameters which will be further discussed in
Sec.~\ref{sec:numerics}.

\section{CP-violating invariants}
\label{sec:invariants}

\par
The generation of a $B-L$ asymmetry from the
decay of the inflaton requires that the theory
contains at least one CP-violating product of
coupling constants which remains unaffected by
field rephasing and is non-real corresponding
to an operator non-invariant under CP
conjugation. Since leptogenesis takes place
at reheating, we work in the vacuum where
$\vev{\phi}=\vev{\bar{\phi}}=M$ and ignore the
$\mu$ term. In previous work \cite{previous},
when calculating the lepton asymmetry from the
CP-violating inflaton decay, we considered only
the couplings of the inflaton originating from
the renormalizable terms in the superpotential
of Eq.~(\ref{W}) (or Eq.~(\ref{WSU2R})). In
fact, when one does also include the couplings
of the inflaton to RHN and Higgs superfields
from the last two explicitly displayed
non-renormalizable terms in the RHS of
Eq.~(\ref{W}) (or Eq.~(\ref{WSU2R})), it turns
out that their contribution to the asymmetry
vanishes exactly. Thus, we will replace these
two superpotential terms by the effective
mass terms $M_i\nu^c_i\nu^c_i$ and couplings
$\zeta_i\nu^c_i(h_1h_2)$ respectively. Since
the only massive particles at reheating are the
inflaton and the RHNs, it is convenient to work
in the basis where the RHN Majorana masses are
diagonal.

\par
In Ref.~\cite{previous}, we considered
rephasing invariants which could survive even
if there were only one matter generation. This
was possible because of the presence of the
${\rm SU}(2)_L$ triplet superfields. In the
current models, we must rely on the presence of
more than one generation. In addition to the
standard leptogenesis CP-violating invariant
composed of four Yukawa coupling constants and
two RHN masses, there is a CP-violating
invariant which can be written in the
``hatted'' basis, where the neutrino Yukawa
couplings are diagonal, as
\beq
I_{0ijk}=\hat{\zeta}_i \hat{\zeta}_j^*
\hat{M}_{\nu^ckj}\hat{M}_{\nu^cki}^*
\eeq
for $i\neq j$ (no summation). However, this
does not survive transformation to the RHN
basis and does not on its own give rise to
diagrams which can create a lepton number
asymmetry in light fields since the
corresponding operator does not involve any
light lepton superfields. Instead, we
``dress'' it with Yukawa coupling constants
to obtain the CP-violating invariant
\beq
I_{ijk} = y_{\nu ij}y_{\nu ik}^*
\zeta_j\zeta_k^* M_j^*M_k
\eeq
for $j\neq k$ (no summation), where we now use
the mass eigenstate basis of
RHNs. When this invariant is used, the
generation of a $B-L$ asymmetry is independent
of the sources of CP violation considered
in previous scenarios and we require novel
decay diagrams. This CP-violating invariant is
minimal in the sense that it involves the
smallest possible number of trilinear
superpotential couplings. We can transform the
first (light lepton) index of this invariant to
the ``hatted'' basis and re-express the
invariant in terms of the ``hatted'' coupling
constants as
\beq
\hat{I}_{ijk}=\sum_{m,n}\hat{y}_i^2 U^c_{ij}
U^{c\ast}_{ik}\hat{\zeta}_m U^c_{mj}
M_j \hat{\zeta}_n^* U^{c\ast}_{nk} M_k,
\eeq
where $\hat{\zeta}_m$ is defined by
\beq
\zeta_n=\sum_m\hat{\zeta}_mU^c_{mn},
\eeq
and we have redefined fields such that $M_i$
and $\hat{y}_i$ are real and positive (the two
last indices of the invariant remain in the RHN
basis). This will be convenient because the
Yukawa couplings in this basis preserve the
individual lepton numbers $L_i$.

\par
Note that the invariant $I_{ijk}$
can be split in two factors $y_{\nu ij}\zeta_j
M_j^*$ and $y_{\nu ik}^*\zeta_k^*M_k$ which
correspond to effective operators with opposite
non-zero $B-L$ charges. These operators involve
only light fields since the heavy ones can be
contracted. One of these two operators includes
bosonic or fermionic $h_1$, $h_2$, while the
other contains their conjugates. These
properties are essential for leptogenesis since
the inflaton field couples at tree level to the
electroweak Higgs superfields $h_1$, $h_2$ and
can decay only to light particles. So, we
conclude that $I_{ijk}$ is, in principle,
suitable for leptogenesis which requires the
interference of two $(B-L)$-violating diagrams
for the inflaton decay. One can further show
that, if the RHN spectrum and couplings are
such that the standard leptogenesis scenario
(without the $\zeta$ couplings) fails, any
invariant which can be useful for leptogenesis
must involve $I_{ijk}$, and thus the effective
coupling constants $\zeta_i$. So, the explicit
violation of ${\rm U}(1)_R$ and matter parity
is essential for our scheme. This is another
novel feature of this scenario.

\section{Baryon asymmetry}
\label{sec:bau}

\par
\begin{figure}[tb]
\centering
\includegraphics[width=\linewidth]{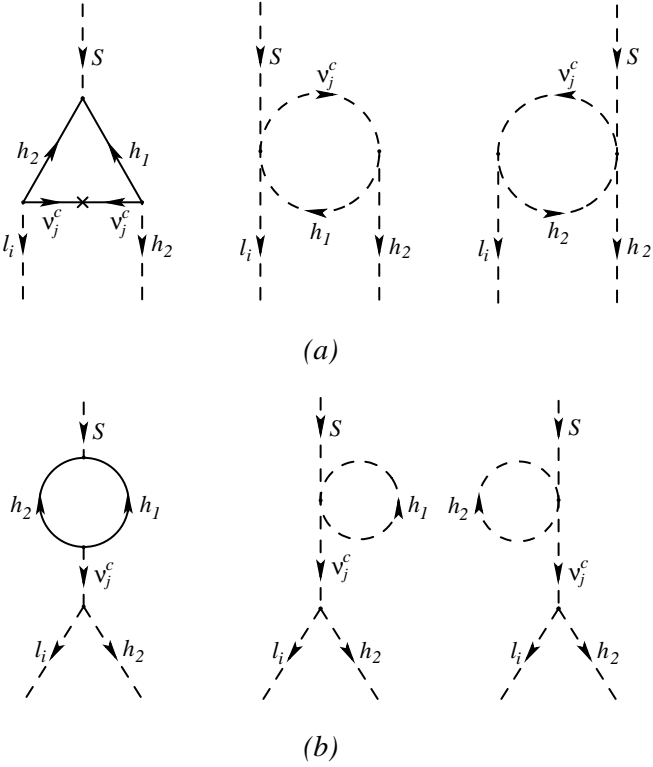}
\caption{The six one-loop
diagrams for the $L_i$-violating decay
$S\rightarrow\tilde{l}_i\,h_2$. The solid
(dashed) lines represent the fermionic
(bosonic) component of the indicated
superfield. The arrows depict the
chirality of the superfields and the
crosses are mass insertions in fermion
lines.}
\label{fig:1}
\end{figure}
\begin{figure}
\centering
\includegraphics[width=\linewidth]{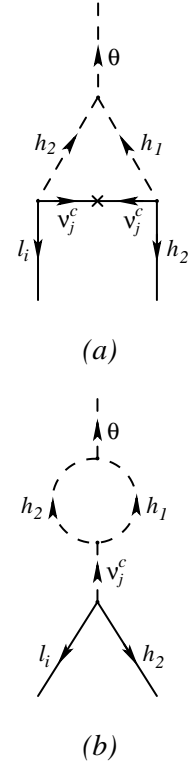}
\caption{ The two one-loop
diagrams for the $L_i$-violating decay
$\theta^*\rightarrow l_i\,\tilde{h}_2$.
The notation is as in Fig.~\ref{fig:1}.}
\label{fig:2}
\end{figure}

In Figs.~\ref{fig:1} and \ref{fig:2}, we
present the diagrams for the $L_i$-violating
decay of $S$ and $\theta^*$ respectively.
Observe that the CP-violating rephasing
invariant $\hat{I}_{ijk}$ corresponds to the
product of coupling constants in the
interference of any
diagram in Fig.~\ref{fig:1} (Fig.~\ref{fig:2})
with any diagram in the same figure where the
exchanged RHN is renamed $\nu^c_k$. This
interference contributes to the $L_i$ asymmetry
due to a partial rate difference in the decays
$S\rightarrow\tilde{l}_i\,h_2$ and $S^*
\rightarrow\tilde{l}_i^*\,h_2^*$ ($\theta^*
\rightarrow l_i\,\tilde{h}_2$ and $\theta
\rightarrow\bar{l}_i\,\bar{\tilde{h}}_2$),
where bar and tilde represent the anti-fermion
and the SUSY partner respectively
\cite{notation}. Both the
decaying inflaton field ($S$ or $\theta^*$),
which is taken at rest, and the decay products
must be on mass shell. For simplicity, we
consider that all the propagating and external
MSSM particles in the diagrams are massless.
Moreover, we perform the calculation in the
limit of exact SUSY.

\par
We also considered diagrams where the inflaton
decays via its couplings to RHN and Higgs
superfields which arise from the superpotential
terms with coefficients $M_{\nu^c ij}$ and
$\lambda_i$. Such diagrams exist only for the
decay of $\theta^*$ and have final states
($l_i\,\bar{\tilde{h}}_1$,
$\bar{l}_i\,\tilde{h}_1$, $\tilde{l}_i\,h_1^*$,
$\tilde{l}_i^*\,h_1$) which are different
from the final state of the diagrams shown in
Fig.~\ref{fig:2}. However, it can be argued
that these diagrams sum to zero in the decay
amplitudes if the final state is
$\tilde{l}_i\,h_1^*$ or $\tilde{l}_i^*\,h_1$.
For the other two cases, the diagrams again sum
to zero but only in the amplitudes-squared
after summing over the phase space of final
particles and in the limit where these
particles are massless.

\par
We will denote by $F^S_{aijn}$ and $F^S_{bijn}$
the ``stripped'' diagrams in Figs.~\ref{fig:1}a
and \ref{fig:1}b respectively with the
dimensionless coupling constants and the
$M_j$ mass insertions factored out (we
keep, though, the $m_{\rm inf}$ factor
appearing in the scalar coupling
$\theta^*h_1h_2$). Here $i$ and $j$ are family
indices and $n=1,2,3$ the serial number of the
diagram. Similarly, the ``stripped'' diagrams
in Figs.~\ref{fig:2}a and \ref{fig:2}b are
$F^\theta_{aij}$ and $F^\theta_{bij}$. In each
case, the contribution to the $L_i$ asymmetry
is proportional to both
$\mbox{Im}\,\hat{I}_{ijk}$
and the imaginary part of the interference of
the relevant ``stripped'' diagrams. Thus, the
total net $L_i$ asymmetries $\eps_{i|S}$ and
$\eps_{i|\theta}$ generated per $S$ and
$\theta^*$ decay respectively are
\bea
\label{epsSTheta}
\eps_{i|S} &=& -\frac{|\lambda|^2}{\Gamma}
\sum_{j,k}\mbox{Im}\,\hat{I}_{ijk}
\sum_{t,t',n,n'}
\mbox{Im}\,[F^S_{tijn}{F^S_{t'ikn'}}^*],
\nonumber \\
\eps_{i|\theta} &=& -\frac{|\lambda|^2}{\Gamma}
\sum_{j,k}\mbox{Im}\,\hat{I}_{ijk}\sum_{t,t'}
\mbox{Im}\,[F^\theta_{tij}{F^\theta_{t'ik}}^*],
\eea
where $\Gamma=|\lambda|^2 m_{\rm inf}/8\pi$
is the common rate of the tree-level decays
$S\rightarrow\bar{\tilde{h}}_1\,
\bar{\tilde{h}}_2$ and $\theta\rightarrow
h_1\,h_2$, the indices $t,t'=a,b$, and
integration over the phase space of the
particles in the final state is implied.

\par
The diagrams are both of the vertex
(Figs.~\ref{fig:1}a and \ref{fig:2}a) and
self-energy (Figs.~\ref{fig:1}b and
\ref{fig:2}b) \cite{covi} type. Each of the
three vertex diagrams in Fig.~\ref{fig:1}a
possesses a logarithmic ultraviolet (UV)
divergence. However, it can be easily shown
that their sum equals $m_{\rm inf}$ times the
vertex diagram in Fig.~\ref{fig:2}a, which is
UV finite. Similarly, one can show that the sum
of the three quadratically divergent
self-energy diagrams in Fig.~\ref{fig:1}b
equals $m_{\rm inf}$ times the self-energy
diagram in Fig.~\ref{fig:2}b. The latter is
though not UV finite. It rather possesses a
logarithmic divergence and, thus, needs
renormalization.

\par
The above relations between the diagrams for
the $L_i$-violating decays of $S$ and
$\theta^*$ imply that
$\eps_{i|S}=\eps_{i|\theta}
\equiv\eps_i$. We thus concentrate on the
calculation of $\eps_{i|\theta}$. The vertex
diagram in Fig.~\ref{fig:2}a being finite (both
its real and imaginary parts) is independent of
the renormalization scheme used. However, the
diagram in Fig.~\ref{fig:2}b involving a
divergent self-energy loop requires us to apply
a renormalization condition. As argued in
Ref.~\cite{pilaftsis}, the appropriate
renormalization scheme, in this case, is the
on-shell (OS) scheme. In a general theory with
scalars $S_i$, the OS conditions on the
renormalized self-energies
$\hat{\Pi}_{ij}(p^2)$ are as follows:
\beq
\mbox{Re}\,\hat{\Pi}_{ij}(\mu_i^2)=
\mbox{Re}\,\hat{\Pi}_{ij}(\mu_j^2)=0
\eeq
for the off-diagonal self-energies
($i\neq j$) and
\beq
\lim_{p^2\rightarrow \mu_i^2}\frac{1}
{p^2-\mu_i^2}\mbox{Re}\,\hat{\Pi}_{ii}(p^2)
=0
\eeq
for the diagonal ones (see {\it e.g.}
Ref.~\cite{Pilaftsisren}). Here we take a basis
where the renormalized mass matrix is diagonal
with eigenvalues $\mu_i$. The imaginary part of
the self-energies is finite and, thus, not
renormalized. In Fig.~\ref{fig:2}b, we have an
off-diagonal self-energy diagram between the
scalars $\theta$ and $\tilde{\nu}^c_j$. Given
that $\theta$ is on mass shell, the real part
of this diagram vanishes in the OS scheme. The
imaginary part, however, gives a finite
contribution.

\par
The asymmetry generated in the lepton number
$L_i$ per decaying inflaton is then
\begin{equation}
\epsilon_i=-m_{\rm inf}^2\sum_{j,k}\mbox{Im}\,
\hat{I}_{ijk}\sum_{t,t'}\mbox{Im}[
{F^t_j}{F^{t'}_k}^*].
\label{ei}
\end{equation}
Here, the indices $t$, $t'$ can be thought of
as indices which refer to the topology of the
diagrams in Fig.~\ref{fig:2}. Namely
$t={\rm trg}$ corresponds to the vertex diagram
in Fig.~\ref{fig:2}a, while $t={\rm self}$ to
the self-energy diagram in Fig.~\ref{fig:2}b.
We find that
\begin{equation}
F^{\rm trg}_j=\frac{1}{16\pi^2m_{\rm inf}^2}
\left[\frac{\pi^2}{6}-{\rm Li}_2\left(1+
\frac{m^2_{\rm inf}}{M^2_j}+i\varepsilon\right)
\right],
\end{equation}
where ${\rm Li}_2$ is the dilogarithm
\cite{dilog}, and
\begin{equation}
F^{\rm self}_j=\frac{1}{4\pi}\frac{i}
{m_{\rm inf}^2-M^2_j}.
\label{Fself}
\end{equation}
This equation for the contribution of the
self-energy diagram holds \cite{pilaftsis}
provided that the decay width of
$\tilde{\nu}^c_j$ is
$\ll |m^2_{\rm inf}-M^2_j|/m_{\rm inf}$, which
is well satisfied in our model if $M_j$ is not
unnaturally close to $m_{\rm inf}$. Also, note
that the contribution to the second sum in the
RHS of Eq.~(\ref{ei}) originating from the
interference of two self-energy diagrams
({\it i.e.} the term with
$t=t'={\rm self}$) vanishes. The reason is that
$F^{\rm self}_j$ is pure imaginary. Finally,
the contributions to the RHS of Eq.~(\ref{ei})
which are diagonal in family space ({\it i.e.}
with $j=k$) also vanish since $\hat{I}_{ijk}$
is real in this case.

\par
The equilibrium conditions including
non-perturbative electroweak reactions, for
temperatures below the mass scale of
superpartners and the critical temperature
$T_c$ of the electroweak phase transition,
yield a relation which allows us to find the
final baryon asymmetry $n_B/s$ in terms of the
$B-3L_i$ asymmetries $n_{B-3L_i}/s$, where
$n_X$ is the density of the quantum number $X$
and $s$ is the entropy density. Note that, in
this regime, all
three quantum numbers $B-3L_i$ are conserved.
This is, however, not true for temperatures
above the scale of sparticle masses, which is
taken to lie above the critical temperature of
the electroweak transition. We assume that at
least one of these quantum numbers, which we
designate as $B-3L_3$, is conserved at all
temperatures after reheating (see
Sec.~\ref{sec:preserve}). Then, in a
temperature range just above the SUSY threshold
where $B-3L_{1,2}$ are violated by processes
involving sparticles, we can determine
$n_{B-3L_{1,2}}/s$ in terms of
$n_{B-3L_3}/s$. Assuming continuity of
$n_{B-3L_i}/s$ as the temperature crosses the
SUSY threshold, we can express $n_B/s$ in terms
of $n_{B-3L_3}/s$ (see Sec.~\ref{sec:chempot}).
The result can be written as
\beq
\frac{n_B}{s}=k\frac{n_{B-3L_3}}{s}.
\eeq
Then if we imagine the inflaton to decay
instantaneously out of equilibrium creating
initial $L_3$ lepton number density
$n_{L_3,{\rm init}}$, we have
\beq
\frac{n_B}{s}=-3k\frac{n_{L_3,{\rm init}}}{s}
=-3k\eps_3\frac{n_{\rm inf}}{s}=-\frac{9k}{4}
\eps_3\frac{T_{\rm reh}}{m_{\rm inf}}
\label{nb}
\eeq
using the standard relation $n_{\rm inf}/
s\equiv (n_S+n_\theta)/s= 3T_{\rm reh}/4
m_{\rm inf}$ for the inflaton number
density. The reheat temperature is given
by
\beq
\label{Trh}
T_{\rm reh}=\left(\frac{45}{2\pi^2 g_\ast}
\right)^{\frac{1}{4}}
(\Gamma m_{\rm P})^{\frac{1}{2}},
\eeq
where $m_{\rm P}\simeq 2.44\times 10^{18}~
{\rm GeV}$ is the reduced Planck scale, and
$g_\ast$ counts the effective number of
relativistic degrees of freedom taking account
of the spin and statistics and is equal to
$228.75$ for the MSSM spectrum.

\section{Effects of R-symmetry violation}
\label{sec:rsym}

\par
We will now explore the possible
phenomenological and cosmological consequences
of the explicit violation of ${\rm U}(1)_R$ in
the superpotential. Our models contain a $Z_2$
matter parity symmetry under which all the
matter (quark and lepton) superfields change
sign. This symmetry is actually the $Z_2$
subgroup of ${\rm U}(1)_R$ and is left unbroken
by the soft SUSY-breaking terms. Matter parity
combined with the $Z_2$ fermion parity under
which all fermions change sign yields R-parity,
which, if unbroken, guarantees the stability of
the LSP. In the present case, however, matter
parity is violated along with the
${\rm U}(1)_R$ by the last explicitly displayed
term in the RHS of Eq.~(\ref{W}) (or
Eq.~(\ref{WSU2R})), which is necessary for
generating the observed BAU. Consequently,
R-parity is explicitly violated and the LSP
can decay rapidly, rendering it unsuitable for
dark matter and leading to distinctive collider
signatures.

\par
The LSP could decay into a pair of electroweak
Higgs bosons and a lepton if it contains a
Higgsino component. The dominant diagrams are
constructed from the ${\rm U}(1)_R$- and
R-parity-violating Yukawa vertices
$\zeta_j\nu^c_j(h_1h_2)$ with the fermionic
$\nu^c_j$ connected to the fermionic $\nu^c_j$
of the Yukawa couplings $y_{\nu ij}(l_ih_2)
\nu^c_j$ via a mass insertion. For the values
of the parameters considered here (see
Sec.~\ref{sec:numerics}), we find that the
resulting LSP life-time can be as low as about
$10^{-1}~{\rm sec}$. However, it is easy to
block kinematically these decay channels of the
LSP by taking its mass to be smaller than twice
the mass of the lightest Higgs boson, which is
very reasonable. Even then, the LSP could decay
to a lepton and an electroweak Higgs boson with
similar life-time. The relevant (one-loop)
diagrams may be obtained by connecting the two
external Higgs lines of the previous diagrams
to a trilinear Higgs vertex. These two-body
decay channels can though also be blocked
kinematically if the LSP mass is taken smaller
than the lightest Higgs boson mass. In this
case, the dominant diagrams for the LSP decay
may be constructed from the above one-loop
diagrams by coupling the external Higgs line to
a pair of fermions (excluding the $t$-quark).
So, we obtain one-loop diagrams leading to
three-body decay of the LSP. They involve an
extra small Yukawa coupling constant which
together with the one-loop factor yields a
suppressed decay rate. The life-time can be of
order $10^{9}~{\rm sec}$ or higher and can be
further enhanced if the LSP is predominantly a
gaugino. Thus, it may be possible to rescue the
LSP as dark matter candidate, especially if its
Higgsino component is suppressed.
Alternatively, if the LSP decays fast, we would
have to ensure that its life-time were less
than $1~{\rm sec}$ to avoid conflict with
primordial nucleosynthesis (see {\it e.g.}
Ref.~\cite{DreinerR}).

\par
Besides the LSP decay, we could also have other
low energy processes which violate R-parity.
The superpotential terms which break R-parity
involve at least one superheavy field. On
integrating out the superheavy fields, we
generically obtain effective R-parity-violating
operators involving only MSSM fields. If these
operators have dimension five or higher, they
do not lead to detectable processes since they
are suppressed by some powers of $M_j$. On the
contrary,  operators with dimension four such
as the effective scalar vertex $h_1h_2h_2^*
\tilde{l}_i^*$ which originates from the
superpotential couplings $\hat{\zeta}_i
\hat{\nu}^c_i(h_1h_2)$, $\hat{y}_i(\hat{l}_ih_2)
\hat{\nu}^c_i$ can lead to low-energy
R-parity-violating processes which may be
detectable in the future colliders (see
Sec.~\ref{sec:numerics}).

\par
As in any leptogenesis scenario with RHNs, we
must ensure \cite{turner} that the primordial
$L_i$ asymmetry (for at least one value of $i$)
is not erased by lepton-number-violating
scattering processes such as $l_i
\tilde{l}_j\rightarrow h_2^*\bar{\tilde{h}}_2$
or $l_i\tilde{h}_2\rightarrow h_2^*
\tilde{l}^*_j$ at all temperatures between
$T_{\rm reh}$ and about $100~{\rm GeV}$. In our
case, due to the presence of the
R-parity-violating superpotential terms, there
exist some extra processes of this type such as
$\tilde{h}_1\tilde{l}_j\rightarrow h_2^*
\bar{\tilde{h}}_2$ or $\tilde{h}_1\tilde{h}_2
\rightarrow h_2^*\tilde{l}^*_j$, which come
from diagrams similar to the ones mentioned
above for the fast LSP decay. In addition to
all these processes which correspond to
effective operators of dimension five (or
higher), we also have dimension-four processes
which violate R-parity (and lepton number) such
as the process
$h_1h_2 \rightarrow h_2\tilde{l}_i$ derived
from the effective four-scalar vertex in the
previous paragraph. We will return to this
issue in the next section.

\par
As already mentioned, the classical flatness of
the inflationary trajectory in the limit of
global SUSY is ensured, in our case, by a
continuous R-symmetry enforcing a linear
dependence of the superpotential on $S$. This
is retained \cite{lss} even after supergravity
corrections, given a reasonable assumption
about the K{\" a}hler potential. The solution
\cite{lr} to the $\mu$ problem is also reliant
on the R-symmetry. These properties are not
affected by the explicit R-symmetry breaking
considered here. Note that, in our scheme, some
R-symmetry-violating couplings are included in
the superpotential and some not. Thanks to the
non-renormalization property of SUSY, this
situation is stable under radiative
corrections, but it may be considered unnatural
since there is no symmetry to forbid the
missing terms.

\section{Preservation of $B-3L_3$}
\label{sec:preserve}

\par
In this section, we will discuss the conditions
for at least one quantum number $B-3L_i$ to be
preserved from reheating all the way down to
the
electroweak phase transition. This is important
since, in the opposite case, the primordial
lepton asymmetry will be erased. Note that the
non-perturbative electroweak sphaleron effects
preserve all these three quantum numbers. So,
violation of some $B-3L_i$ can only take place
if the corresponding $L_i$ is violated
perturbatively. (Recall that $B$ is conserved
to all orders in perturbation theory and the
non-perturbative QCD instanton interactions
conserve $B$ and all the $L_i$'s.)

\par
Dangerous $L_i$-violating reactions that could
wash out any primordial asymmetry in $B-3L_i$
include dimension-four scalar interactions
arising from the F-term
$|F_{\hat{\nu}^c_k}|^2$, which violate $L_i$
either through the R-parity-violating operator
$\hat{\zeta}_k\hat{\nu}^c_k
(h_1h_2)$ or if they involve two Yukawa
couplings $\hat{y}_{\nu ik}(\hat{l}_ih_2)
\hat{\nu}^c_k$ and $\hat{y}_{\nu jk}(\hat{l}_j
h_2)\hat{\nu}^c_k$ with $i\neq j$. Note that
the latter possibility exists if $y_\nu$ is not
simultaneously diagonalizable with $y_e$ (the
$3\times 3$ matrix with elements $y_{eij}$). In
the case where the lightest RHN ($\nu^c_1$) has
a mass less than $T_{\rm reh}$, direct thermal
production of $\nu^c_1$ also takes place. We
then have an extra source of $L_i$ violation from
the Yukawa coupling of $\nu^c_1$ to $\hat{l}_i$.
Finally, we must consider $L_i$-violating
effective dimension-five operators arising from
fermionic $\nu^c$ exchange with mass insertion
(and also without mass insertion in the case
$T_{\rm reh}> M_1$).

\subsection{Dimension-four operators}

\par
As mentioned earlier, individual lepton numbers
$L_i$ for the MSSM superfields can only be
defined in the ``hatted'' basis where the
charged lepton Yukawa coupling constant matrix
$\hat{y}_e$ is diagonal. In the
${\rm SU}(2)_R$-symmetric case, $y_\nu$ can be
diagonalized simultaneously with $y_e$. In the
non-${\rm SU}(2)_R$-symmetric case, however,
this is in general not possible. It is though
possible to strongly restrict the form of
$\hat{y}_\nu$ by using the remaining freedom of
performing unitary rotations in the flavor
space of the RHN superfields and applying the
condition that $L_i$ (for some particular value
of $i$) is conserved by dimension-four
operators. We start by considering the
four-scalar operator $\tilde{l}_ih_2
\tilde{l}_j^*h_2^*$ with coupling constant
$\hat{y}_{\nu ik}\hat{y}_{\nu jk}^*$ which, for
$j\neq i$, violates $L_i$. In order to retain
the $L_i$ asymmetry, we must impose the
condition
\beq
\sum_k\hat{y}_{\nu ik}\hat{y}_{\nu jk}^*=0
\label{yy*}
\eeq
for $j\neq i$. For definiteness, let us choose
to preserve the $L_3$ asymmetry, thus $i=3$. An
appropriate rotation in the $\nu^c_k$ space can
bring the 3-vector $\hat{y}_{\nu 3k}$ on the
$\nu^c_3$ axis, {\it i.e.} can make
$\hat{y}_{\nu 31}$, $\hat{y}_{\nu 32}$ to
vanish. Equation~(\ref{yy*}) then implies that
the 3-vectors $\hat{y}_{\nu 1k}$ and
$\hat{y}_{\nu 2k}$ lie in the $\nu^c_1-\nu^c_2$
subspace, {\it i.e.} $\hat{y}_{\nu 13}=
\hat{y}_{\nu 23}=0$. Using the remaining
freedom of rotations in this subspace, we can
bring $\hat{y}_{\nu 2k}$ on the $\nu^c_2$ axis,
{\it i.e.} make $\hat{y}_{\nu 21}=0$. So the
only non-vanishing $\hat{y}_{\nu ik}$'s are the
diagonal ones and $\hat{y}_{\nu 12}$. The
diagonal elements $\hat{y}_i$ of $\hat{y}_\nu$
can be further made real and positive by
absorbing their complex phases into the RHN
superfields, while $\hat{y}_{\nu 12}$ remains
arbitrary and complex. For simplicity, we have
chosen $\hat{y}_{\nu 12}=0$ so that $y_\nu$ is
simultaneously diagonalizable with $y_e$. We
still have to consider the four-scalar operator
$h_1h_2 h_2^*\tilde{l}_3^*$ with coupling
constant $\hat{\zeta}_3\hat{y}_3^*$ which also
violates $L_3$. Therefore, in order to retain
the $L_3$ asymmetry, we must further impose the
condition
\beq
\hat{\zeta}_3=0.
\label{noerase}
\eeq
Note that it is by no means necessary for this
equation, or the other constraints on
$L_3$-violating couplings, to be satisfied
exactly. The dangerous dimension-four operators
will not come into thermal equilibrium if their
effective coupling constants are less than
about $10^{-7}$. However, we take them to
vanish exactly in order to simplify the
discussion.

\subsection{Dimension-five operators}

In Sec.~\ref{sec:chempot}, we will assume that,
for a temperature range just above the scale of
the superpartner masses, the $L_1$ and $L_2$
quantum numbers are violated due to the
four-scalar operators involving $\zeta$
couplings, whereas $L_3$ is perturbatively
conserved for all temperatures after reheating.
The latter is guaranteed for dimension-four
operators by Eq.~(\ref{noerase}), but needs to
be checked for effective dimension-five
operators with $\nu^c$ exchange, since the RHN
mass terms explicitly violate the lepton
numbers $L_i$. We consider these operators
first for the case without ${\rm SU}(2)_R$
symmetry where the RHN masses are all at or
above the inflaton mass scale (see
Sec.~\ref{sec:numerics}), and then for the case
of ${\rm SU}(2)_R$ symmetry. In both cases, we
estimate the rate of reaction mediated by a
chirality-conserving RHN exchange as
\beq
\Gamma\sim T^3\left|\sum_k\frac{A_kT}
{\max(M_k^2,T^2)}\right|^2,
\label{Gnoflip}
\eeq
where $A_k$ is the product of the dimensionless
coupling constants which corresponds to the
diagram with an exchange of a RHN with mass
$M_k$. In the case of a chirality-changing
transition with mass insertion on the $\nu^c$
propagator, the relevant formula is
\beq
\Gamma\sim T^3\left|\sum_k\frac{A_kM_k}
{\max(M_k^2,T^2)}\right|^2.
\label{Gflip}
\eeq

\subsubsection{Non-${\rm SU}(2)_R$-symmetric
case}
\label{nonSU2R}

\par
When all RHNs have masses of the order of
$m_{\rm inf}$ or greater, we have $M_i>T$.
Equation (\ref{Gflip}) then becomes
\beq
\Gamma\sim |\mathcal{A}|^2 T^3,
\label{rate}
\eeq
where $\mathcal{A}=\sum_kA_k/M_k$. For
scattering processes of the type $l_ih_2
\rightarrow \bar{l}_jh_2^*$, which are
mediated by a RHN exchange with mass insertion,
this is
\beq
\mathcal{A}_{ij}=-\hat{y}_{i}
(\hat{M}_{\nu^c}^{-1})_{ij}\hat{y}_{j}\simeq
\frac{2}{v^2} \hat{m}_{ij},
\label{scriptA}
\eeq
where, as explained above, we take the neutrino
Yukawa couplings to be diagonal in the charged
lepton flavor basis (``hatted'' basis) and also
approximate the up-type Higgs VEV to $v\simeq
174~{\rm GeV}$ (large $\tan\beta$ limit). The
light neutrino mass matrix in the charged
lepton basis is found as $\hat{m}=V^*\bar{m}
V^\dag$ for given values of the MNS matrix $V$
and light neutrino masses $m_i$. Hence, we find
that $\mathcal{A}_{ij}$ has entries of at most
$10^{-15}~{\rm GeV}^{-1}$, assuming a
hierarchical spectrum of light neutrinos
\cite{Glashow}. The ratio of the rate of the
$L_3$-violating scattering processes involving
two Yukawa couplings and a chirality-changing
RHN exchange to the Hubble rate is thus
\beq
\frac{\Gamma}{H}\approx\frac{m_{\rm P}T
|\mathcal{A}_{3j}|^2}{5}\approx T
|\mathcal{A}_{3j}|^2 \times 4.9\times 10^{17}
~\mbox{GeV}.
\label{GammaonH}
\eeq
Therefore, even for reheat temperatures which
are as high as $3\times 10^{10}~{\rm GeV}$,
these scattering processes are well out of
equilibrium after reheating.

\par
We also have to take into account scattering
processes of the type $l_3h_2\rightarrow
{\bar{\tilde h}}_1h_2^*$  with one Yukawa and
one $\hat{\zeta}_j$ coupling, $j\neq 3$,
mediated by a chirality-changing RHN exchange.
The amplitude for these processes is
\beq
\sum_{j=1,2}\mathcal{A}_{3j}~\frac
{\hat{\zeta}_j}{\hat{y}_j}.
\eeq
For given values of the light neutrino
parameters, $\hat{\zeta}_j$ and
$\hat{y}_j$, the rate may be found. The worst
case occurs in the small $m_1$ limit for which
the rate of these scattering processes at
$T=3\times 10^{10}~{\rm GeV}$ is estimated as
\beq
\frac{\Gamma}{H} \approx 3.4\left|\frac{
\hat{\zeta}}{(m^D_1/{\rm GeV})}\right|^2+
190\left|\frac{
\hat{\zeta}}{(m^D_2/{\rm GeV})}\right|^2,
\eeq
where $m^D_i$ are the (diagonal) Dirac neutrino
masses and we take $\hat{\zeta}_1=i
\hat{\zeta}_2$ with $|\hat{\zeta}_1|=
|\hat{\zeta}_2|\equiv |\hat{\zeta}|$ (see
Sec.~\ref{sec:numerics}). Hence $L_3$ is safely
preserved after reheating if we have
\beq
\frac{m^D_1}{\rm GeV}\gtrsim 1.8|\hat{\zeta}|,
\ \
\frac{m^D_2}{\rm GeV}\gtrsim 14|\hat{\zeta}|,
\eeq
which are easily satisfied within the region of
parameter space where the model constraints are
fulfilled. Note that dimension-five operators
mediated by a chirality-conserving RHN exchange
are even more suppressed as one can easily
deduce by comparing Eqs.~(\ref{Gnoflip}) and
(\ref{Gflip}). So, no $L_3$ violation by
dimension-five operators is encountered in the
non-${\rm SU}(2)_R$-symmetric case after
reheating.

\subsubsection{${\rm SU}(2)_R$-symmetric case}

\par
In the ${\rm SU}(2)_R$-symmetric case, the
neutrino Yukawa coupling constants in the
``hatted'' basis are unambiguously determined
to be given by the asymptotic relation
$\hat{y}_{\nu}=(1/v\cos\beta)~{\rm diag}
(m_e,\,m_\mu,\,m_\tau)$. For the
purposes of calculating thermal effects near
the reheat temperature, we can use the
asymptotic values of Yukawa coupling constants
since these constants do not run very much
between the GUT scale and $T_{\rm reh}$. Our
best-fit value of $\tan\beta$ is around 55,
thus the numerical values of the $\hat{y}_i$'s
are given by
$\hat{y}_{\nu}={\rm diag}(0.00014,0.028,0.66)$
approximately, taking into account the RG
evolution of the charged lepton Yukawa coupling
constants to the GUT scale. The $\nu^c$ masses
are also determined uniquely by the light
neutrino masses and the MNS matrix. As
mentioned, the RHN masses must satisfy
$M_3\lesssim M^2/M_S$ and $M_2>
m_{\rm inf}/2$. The lightest RHN mass ($M_1$)
turns out to be very small (of order $10^7~
{\rm GeV}$) in this case (see
Sec.~\ref{sec:numerics}). However, for our
chosen parameter values, these bounds may be
satisfied with a similar choice of light
neutrino parameters to that for the
non-${\rm SU}(2)_R$-symmetric case. Hence the
entries in the matrix $\mathcal{A}$ defined in
Eq.~(\ref{scriptA}) are of similar size. Thus,
for temperatures that are smaller than all the
RHN masses, the $L_3$-violating dimension-five
operators involving two Yukawa couplings and a
chirality-changing RHN exchange are out of
equilibrium.

\par
However, we also have to take into account the
fact that we have a range of temperatures
which are higher than the lightest RHN mass
eigenstate $M_1$. In
this case, the contribution of $\nu^c_1$ to
scattering amplitudes will be different from
its contribution in the previous case.
Considering the propagator for $\nu^c_1$
exchange with mass insertion, instead of the
factor $1/M_1$ assumed in Eq.~(\ref{scriptA}),
we obtain $M_1/T^2$. But this will evidently
give a smaller rate than the estimate in
Eq.(\ref{rate}) (which is incorrect in this
case) if there are no cancellations between
contributions of different RHNs (cancellations
appear very unlikely given the large hierarchy
in the $\nu^c$ masses, and this can be checked
numerically). Thus, the $L_3$-violating
scattering rate, in this case, is always equal
to or smaller than the estimate in
Eq.~(\ref{GammaonH}), which is still smaller
than $H$. Consequently, the $L_3$-violating
scattering processes from
dimension-five operators which involve two
Yukawa couplings and a chirality-changing RHN
exchange are expected to be out of equilibrium
at all temperatures after reheating in the
${\rm SU}(2)_R$-symmetric case too.

\par
We also have scattering through $\nu^c$
exchange without mass insertion involving two
Yukawa couplings. This does not violate the
total lepton number, but could convert $L_3$ to
$L_j$ ($j\neq 3$). The $\nu^c_1$ propagator now
varies as $1/T$ for $T>M_1$ or $T/M_1^2$ for
$T<M_1$, which is, in both cases, suppressed
relative to the dependence of $1/M_1$ in
Eq.~(\ref{scriptA}). The situation is similar
in the case of $\nu^c_{2,3}$ exchange where the
discussion at the end of Sec.~\ref{nonSU2R}
applies. Hence the scattering rate is
suppressed relative to the rate in
Eq.~(\ref{GammaonH}), and is again smaller than
the Hubble rate for temperatures below
$3\times 10^{10}~{\rm GeV}$. One can further
show that this conclusion remains valid  even
if we replace one of the Yukawa couplings by a
$\hat{\zeta}$ coupling in all cases (with and
without mass insertion).

\par
Since we are considering temperatures above
$M_1$, there will be a thermal density of
$\nu^c_1$ particles and $L_3$ may be violated,
say, by direct scattering of $l_3$ and $h_2$
into $\bar{\nu}^c_1$ (see also
Ref.~\cite{DreinerR}), where the RHN will
subsequently decay mainly to $l_1h_2$ (or
$\tilde{l}_1h_2^*$). The rate for this is
similar to the rate for any dimension-five
process through $\nu^c_1$
exchange without mass insertion (an equivalent
process if $\nu^c_1$ is put on mass shell),
except that the coupling constant is simply
$\hat{y}_3U^c_{31}$. The relevant bound is now
\beq
|\hat{y}_3U^c_{31}|^2 \times 4.9\times
10^{17}\,{\rm GeV}\lesssim T,
\label{nuc1}
\eeq
which, for $T\approx 10^{7}~{\rm GeV}$, implies
\beq
|U^c_{31}|\lesssim 10^{-5}.
\label{Uc31}
\eeq
This bound may in fact be respected for some
parameter choices, but we will see that the
${\rm SU}(2)_R$-symmetric case is ruled
out by other considerations, namely the
smallness of off-diagonal elements of $U^c$.

\section{Chemical potentials and $n_B/s$}
\label{sec:chempot}

\par
In order to find the final baryon asymmetry, we
consider the evolution of the net number
densities of all the relativistic species
between the time of reheating and the time at
which all $B$-violating interactions are out of
equilibrium and the baryon number is
effectively conserved. At any given
temperature, we determine which reactions
(including the non-perturbative ones,
{\it i.e.} the sphaleron or QCD instanton
effective vertices) are in equilibrium. This is
the case if the rate of reaction normalized to
the particle number density is greater than the
Hubble rate. Then one can deduce which additive
quantum numbers are conserved by these
reactions. The equilibrium number density $n$
of a relativistic species minus the number
density $\bar{n}$ of its antiparticle is
given, in terms of its chemical potential
$\mu$, by the standard formula
\beq
\frac{n-\bar{n}}{s}=\frac{15g}{4\pi^2g_\ast}
\left\{\ba{l}2\
\mbox{(bosons)} \\ 1\
\mbox{(fermions)} \ea \right\}
\frac{\mu}{T},
\eeq
where $g$ is the number of internal degrees of
freedom of the species. For each reaction in
equilibrium, the sum of chemical potentials of
the reaction products is equal to the sum of
chemical potentials of the reactants. Using the
equilibrium equations, one can express
\cite{turner,IbanezQ} the net number density of
any species as well as the density of any
quantum number in terms of a small set of
chemical potentials. Eliminating these chemical
potentials, one can determine the densities of
all quantum numbers given the densities of the
conserved ones. When some reactions go out of
equilibrium, one may find the new densities of
the quantum numbers which cease to be violated
by imposing that the densities remain
unchanged. We use the convention \cite{IbanezQ}
that the same symbol denotes the species and
its chemical potential. For fermions, we count
each left-handed species separately, hence a
non-zero chemical potential for the gluino
$\tilde{g}$ (say) implies an asymmetry between
left- and right-handed gluinos.

\par
We will first perform the chemical potential
analysis in the period which follows reheating
and for temperatures higher than the mass scale
of the superpartners and the critical
temperature $T_c$ for the electroweak phase
transition. In this regime, the SM gauge group
is unbroken and we will assume that the
spectrum coincides with the MSSM one. This is
certainly the case in our model provided that
all the RHN masses are larger than
$T_{\rm reh}$. In the ${\rm SU}(2)_R$-symmetric
case, however, this analysis will only hold for
temperatures which are lower than the lightest
RHN mass, which is smaller than $T_{\rm reh}$.
Gauge interactions imply \cite{turner} that the
gluons, the photon and the $B$ gauge boson have
vanishing chemical potential at all times, as
do the $W^{\pm}$ bosons above the electroweak
phase transition. The interactions of squarks
with all three types of SM gauge bosons and the
corresponding gauginos also ensure
\cite{IbanezQ} that all gauginos have the same
chemical potential, which we denote by
$\tilde{g}$.

\par
Yukawa interactions originating from the
superpotential are found to be in equilibrium
above the electroweak scale, implying the
relations
\bea
u^c&=&-q-h_2,
\nonumber \\
d^c&=&-q-h_1,
\nonumber \\
e^c_i&=&-l_i-h_1.
\eea
Due to the Cabbibo-Kobayashi-Maskawa
mixing, the quark chemical potentials are
independent of the generation number. In the
leptonic sector, the situation is different:
due to the very small scale of masses,
neutrino oscillations are not \cite{DreinerR}
in equilibrium at the temperatures of interest;
and, as has been shown, interactions mediated
by massive RHN exchange may not be either. Thus
the charged lepton and neutrino chemical
potentials may differ between the three
generations in the ``hatted'' basis, where the
lepton Yukawa couplings are diagonal. Note
that, although the chemical potentials $l_i$
(or $\nu_i$, $e_i$) and $e^c_i$ are,
throughout, in the ``hatted'' basis, we
suppress the hats for simplicity of notation.

\par
SUSY-breaking soft terms mediate transitions
which are \cite{IbanezQ} in equilibrium for
temperatures below about $10^7~{\rm GeV}$ and
above the scale of soft scalar masses (for
simplicity, we will assume a common SUSY
threshold). In this regime, the gaugino
chemical potential vanishes and all members of
a supermultiplet have equal chemical potential.
At higher temperatures, we have the following
relations \cite{error} between the components
of the chiral superfields:
\bea
\tilde{q}&=&q+\tilde{g},
\nonumber \\
\tilde{u}^c&=&u^c+\tilde{g},
\nonumber \\
\tilde{d}^c&=&d^c+\tilde{g},
\nonumber \\
\tilde{l}_i&=&l_i+\tilde{g},
\nonumber \\
\tilde{e}^c&=&e^c+\tilde{g},
\nonumber \\
\tilde{h}_{1,2}&=&h_{1,2}-\tilde{g}.
\eea
Then the baryon and lepton number densities are
\bea
\frac{n_B}{s} &=& \frac{15}{4\pi^2g_\ast T}~3N_g
\left(4q+h_1+h_2\right),
\nonumber \\
\frac{n_{L_i}}{s} &=& \frac{15}{4\pi^2g_\ast T}
~[3(3l_i+h_1)+2\tilde{g}],
\eea
where $N_g$ is the number of generations.
The condition that the electric charge density
should vanish is
\beq
\frac{n_Q}{s}=\frac{15}{4\pi^2g_\ast T}~3[2N_gq-
2\sum_il_i+(2N_g+1)(h_2-h_1)]= 0.
\label{nQ}
\eeq

\par
We must now also consider the ${\rm SU}(2)_L$
and ${\rm SU}(3)_c$ non-perturbative
interactions ('t Hooft vertices), which involve
all the left-handed fermions transforming
non-trivially under the corresponding group and
are unsuppressed at these temperatures. They
give the following relations:
\bea
\mbox{SU}(2)_L&:& 3N_g q+\sum_il_i+h_1+
h_2+2\tilde{g}=0,
\nonumber \\
\mbox{SU}(3)_c&:& N_g (h_1+h_2)-6\tilde{g}=0,
\eea
which together with Eq.~(\ref{nQ}) yield
\bea
q &=& -\frac{1}{3N_g}\sum_il_i-
\frac{2(N_g+3)}{3N_g^2}~\tilde{g},
\nonumber \\
h_1 &=& -\frac{4}{3(2N_g+1)}\sum_il_i+
\frac{16N_g+3}{3N_g(2N_g+1)}~\tilde{g},
\nonumber \\
h_2 &=& \frac{4}{3(2N_g+1)}\sum_i l_i+
\frac{20N_g+15}{3N_g(2N_g+1)}~\tilde{g}.
\eea
So, we are left only with four independent
chemical potentials ($l_i$, $\tilde{g}$), in
terms of which the baryon and lepton number
densities are given by
\bea
\frac{n_B}{s} &=& \frac{15}{4\pi^2g_\ast T}
\left(-4\sum_il_i+\frac{10N_g-24}{N_g}~
\tilde{g}\right),
\nonumber \\
\frac{n_{L_i}}{s} &=& \frac{15}{4\pi^2g_\ast T}
\Biggl(9l_i-\frac{4}{2N_g+1}\sum_jl_j
\nonumber \\
& & +\frac{4N_g^2+18N_g+3}{N_g(2N_g+1)}~
\tilde{g}\Biggr),
\label{nbnli}
\eea
which hold regardless of the details of the
model.

\par
The scalar potential term $|\partial W/\partial
\hat{\nu}^c_i|^2$ gives rise to operators
mediating the transition $h_1 h_2
\leftrightarrow h_2\tilde{l}_{i}$ for $i=1$,
$2$ (recall that we imposed the condition that
$\hat{\zeta}_3=0$). These reactions are in
equilibrium just above the superpartner
threshold and imply that $l_{1,2}=-\tilde{g}
+h_1$. Moreover, as already mentioned, below
a temperature $10^7~{\rm GeV}$ and above
the superpartner mass threshold, $\tilde{g}$
vanishes. So, just above the SUSY threshold,
we have only one independent chemical
potential (say $l_3$) and Eq.~(\ref{nbnli})
yields
\bea
\frac{n_B}{s}&=&\frac{15}{4\pi^2g_\ast T}
\frac{-12(2N_g+1)}{6N_g+11}~l_3,
\nonumber \\
\frac{n_{L_{1,2}}}{s}&=&\frac{15}
{4\pi^2g_\ast T}
~\frac{-48}{6N_g+11}~l_3,
\nonumber \\
\frac{n_{L_3}}{s}&=&\frac{15}{4\pi^2g_\ast T}
~\frac{3(18N_g+29)}{6N_g+11}~l_3.
\eea
Eliminating $l_3$, we can express all the
quantum numbers in terms of one of them. In
particular, we find that
\bea
\frac{n_{B}}{s}&=&\frac{4(2N_g+1)}{6N_g+11}~
\frac{n_{B-3L_3}}{s},
\label{nb-3l3}
\\
\frac{n_{B-3L_{1,2}}}{s}&=&\frac{4(2N_g-11)}
{62N_g+91}~\frac{n_{B-3L_3}}{s},
\label{nb-3l12}
\eea
which also gives
\beq
\sum_i\frac{n_{B-3L_i}}{s}=\frac{3(26N_g+1)}
{62N_g+91}~\frac{n_{B-3L_3}}{s}.
\label{snb-3li}
\eeq
Note that $B-3L_3$ is the only conserved
quantum number just above the SUSY threshold.
It is actually the only quantum number which
remains conserved at all times after reheating
and, thus, the corresponding asymmetry is equal
to $-3$ times $n_{L_3,{\rm init}}/s$.
Equation~(\ref{nb-3l3}) then yields
\beq
\frac{n_{B}}{s}=-\frac{12(2N_g+1)}{62N_g+91}
\frac{n_{L_3,{\rm init}}}{s}.
\eeq
This gives the correct result for the final
baryon number density provided that $i$) the
electroweak phase transition is strongly
first order so that no $B$-violating
${\rm SU}(2)_L$ sphaleron interactions are in
equilibrium in the broken phase, and
$ii$) the scale of the superpartner masses is
smaller than the critical temperature $T_c$,
{\it i.e.} the final value of $n_B/s$ is fixed
at a temperature at which reactions involving
superpartners are in equilibrium.

\par
However, if superpartners decouple while
sphalerons are still active, one must consider
the equilibrium equations without the
superpartner contributions, and perform a
non-SUSY analysis \cite{turner}. In this case,
we have two possibilities: either the
electroweak phase transition is strongly first
order ({\it i.e.} the sphalerons freeze
out very quickly after the transition) and the
final baryon number is fixed in the unbroken
phase, or it is second order (or weakly first
order) and the sphalerons continue to be in
equilibrium in the broken phase. Recent data
indicate that the weakly first-order transition
is more likely (see {\it e.g.}
Ref.~\cite{bcw}). We will thus analyze this
case in detail.

\par
With superpartners and heavy Higgs bosons
decoupled, we are left \cite{turner} with the
following chemical potentials which {\em a
priori} may be non-zero: $u$, $d$
(corresponding, respectively, to the up- and
down-type component of the left-handed
${\rm SU}(2)_L$ doublet $q$), $u^c$, $d^c$,
$e_i$, $e^c_i$, $\nu_i$, $h$ (corresponding to
the lightest Higgs boson), and $W^-$. For
$T>T_c$, the chemical potential of the $W^-$
gauge boson vanishes, while below $T_c$, the
$W^-$ boson may have a non-zero chemical
potential, but $h=0$ \cite{turner,goldstone}.
This immediately fixes the ${\rm SU}(2)_L$
singlet chemical potentials as $u^c=-u$,
$d^c=-d$, $e^c_i=-e_i$. Gauge interactions
imply $d=u+W^-$, $e_i=\nu_i+W^-$, leaving us,
for $T<T_c$, with $N_g+2$ independent chemical
potentials $u$, $\nu_i$, $W^-$. The condition
that the electric charge vanish leads to the
relation
\beq
N_g u-\sum_i\nu_i-(2N_g+3) W^-=0,
\label{W-}
\eeq
which determines $W^-$ in terms of $u$ and
$\nu_i$. The baryon and lepton number densities
can be easily calculated and, after eliminating
$W^-$ by using Eq.~(\ref{W-}), are given by
\bea
\frac{n_B}{s} &=& \frac{15}{4\pi^2g_\ast T}
\frac{2N_g}{2N_g+3}\left( (5N_g+6)u-\sum_i
\nu_i\right),
\nonumber \\
\frac{n_{L_i}}{s} &=& \frac{15}{4\pi^2g_\ast T}
\left(3\nu_i+2\frac{N_g u-\sum_j\nu_j}{2N_g+3}
\right).
\label{nbnlsm}
\eea
Now consider the ${\rm SU}(2)_L$ sphaleron
interaction. Its equilibrium condition implies
that $N_g(u+2d)+\sum_i\nu_i$ should vanish,
which fixes $u$ in terms of $\nu_i$:
\beq
u=-\frac{3\sum_i\nu_i}{N_g(8N_g+9)}.
\eeq
Equation~(\ref{nbnlsm}) then yields
\bea
\frac{n_B}{s} &=& -\frac{15}{4\pi^2g_\ast T}
\frac{4(2N_g+3)}{8N_g+9}\sum_i\nu_i,
\\
\frac{n_{B-3L_i}}{s} &=& -\frac{15}
{4\pi^2g_\ast T}\left(9\nu_i+\frac{4(2N_g-3)}
{8N_g+9}\sum_j\nu_j\right).
\nonumber
\eea
The three quantum numbers $B-3L_i$ are
separately conserved after superpartners
decouple, because reactions violating $L_{1,2}$
are out of equilibrium. We solve for $\nu_i$ in
terms of $n_{B-3L_i}$ and substitute back into
the baryon number density to obtain
\beq
\frac{n_B}{s}=\frac{4(2N_g+3)}{3(32N_g+15)}
\sum_i \frac{n_{B-3L_i}}{s}.
\label{lowenergy}
\eeq
From continuity, the constant values of the
asymmetries $n_{B-3L_i}/s$ for temperatures
lower than the SUSY threshold should be
identical to the values of these asymmetries
just above this threshold where the
superpartners are still in equilibrium and
$\tilde{g}=0$. In this regime, $B-3L_{1,2}$ are
violated and the values of the corresponding
asymmetries are expressed in terms of the
conserved $B-3L_3$ asymmetry. Assuming that the
SUSY threshold lies higher than $T_c$, the
$B-3L_{1,2}$ asymmetries are given by
Eq.~(\ref{nb-3l12}). Then, applying
Eqs.~(\ref{snb-3li}) and (\ref{lowenergy}), we
find that the final baryon number density is
given by
\bea
\frac{n_B}{s} &=&
\frac{4(26N_g+1)(2N_g+3)}{(62N_g+91)(32N_g+15)}
\frac{n_{B-3L_3}}{s}
\nonumber \\
&=&-\frac{12(26N_g+1)(2N_g+3)}
{(62N_g+91)(32N_g+15)}
\frac{n_{L_3,{\rm init}}}{s}
\nonumber \\
&=&-\frac{2844}{10249}
\frac{n_{L_3,{\rm init}}}{s}
\eea
for $N_g=3$.

\section{Numerical results}
\label{sec:numerics}

\par
We make the assumption that the LSPs are
long-lived and their thermal production is
negligible, which holds in many cases. So, the
LSPs which possibly contribute to the cold dark
matter (CDM) in the present universe can
originate solely from the late decay of the
gravitinos which were thermally produced at
reheating. Their relic abundance is \cite{km}
\beq
\Omega_{\rm LSP}h^2=0.037\left(
\frac{m_{\rm LSP}}{100~{\rm GeV}}\right)\left(
\frac{T_{\rm reh}}{10^{10}~{\rm GeV}}\right),
\label{lsp}
\eeq
where $\Omega_{\rm LSP}$ is the relic density
of the LSPs in units of the critical density of
the universe, $h$ is the present Hubble
parameter in units of
$100~{\rm km}~{\rm sec}^{-1}~{\rm Mpc}^{-1}$,
and $m_{\rm LSP}$ is the LSP
mass. We further assume that the CDM in the
universe consists exclusively of LSPs. Taking
the best-fit value for the CDM abundance from
the WMAP data \cite{wmap}, {\it i.e.}
$\Omega_{\rm CDM}h^2\simeq 0.1126$, and putting
$m_{\rm LSP}=100~{\rm GeV}$, Eq.~(\ref{lsp})
yields $T_{\rm reh}\simeq 3.04\times 10^{10}~
{\rm GeV}$. This high value of $T_{\rm reh}$,
chosen here to maximize the BAU, can well be
\cite{gravitino} compatible with the gravitino
constraint \cite{khlopov,gravitino} provided
that the branching ratio of the gravitino decay
to a photon and a photino is adequately small.

\par
The R-symmetry-violating effective coupling
constants $\hat{\zeta}_1$, $\hat{\zeta}_2$ are
taken to satisfy the relation $\hat{\zeta}_1=
i\hat{\zeta}_2$, {\it i.e.} the CP-violating
relative phase between them is taken equal to
$\pi/2$. The magnitude of these coupling
constants is maximized, {\it i.e.}
$|\hat{\zeta}_1|=|\hat{\zeta}_2|=M/M_S$. These
choices maximize the generated baryon
asymmetry. In order to preserve perturbativity
and to have a good effective field theory, we
require $M\lesssim 0.1M_S$.

\par
As already mentioned, the light neutrino
mass-squared differences and mixing angles are
generally taken to lie in the $2\sigma$
confidence intervals in Ref.~\cite{maltoni}. In
the non-${\rm SU}(2)_R$-symmetric case, in
particular, we take them to coincide with
their best-fit values in Ref.~\cite{maltoni}:
$\delta m_{21}^2=8.1\times 10^{-5}~{\rm eV}^2$,
$\delta m_{31}^2=2.2\times 10^{-3}~{\rm eV}^2$,
$\sin^2\theta_1=0.5$, $\sin^2\theta_2=0$, and
$\sin^2\theta_3=0.3$. For simplicity, in the
MNS matrix $V$, we set the Dirac phase $\delta$
and the Majorana phases $\alpha$ and $\beta$ to
zero. This highlights the fact that our
mechanism works in the absence of any
CP-violating phases in the standard
(R-parity-conserving) couplings.

\subsection{Non-${\rm SU}(2)_R$-symmetric case}
\label{sec:numernonSU2R}

\par
We will first examine the
non-${\rm SU}(2)_R$-symmetric case. We fix the
parameter $\kappa$ to the value $10^{-4}$. The
cosmic microwave background explorer (COBE)
value of the quadrupole anisotropy of the CMBR
($(\delta T/T)_Q\simeq 6.6\times 10^{-6}$)
\cite{cobe} is then reproduced for $\lambda
\simeq 3.48\times 10^{-4}$ ($>\kappa$ as it
should) and $M\simeq 5.63\times 10^{15}~
{\rm GeV}$. Thus, $m_{\rm inf}\simeq 7.96
\times 10^{11}~{\rm GeV}$. Note that the values
of $\lambda$ and $m_{\rm inf}$ are consistent
with the value of $T_{\rm reh}$ (see
Eq.~(\ref{Trh})). The spectral index of density
perturbations comes out practically equal to
unity.

\begin{figure*}[t]
\centering
\includegraphics[width=135mm,angle=0]
{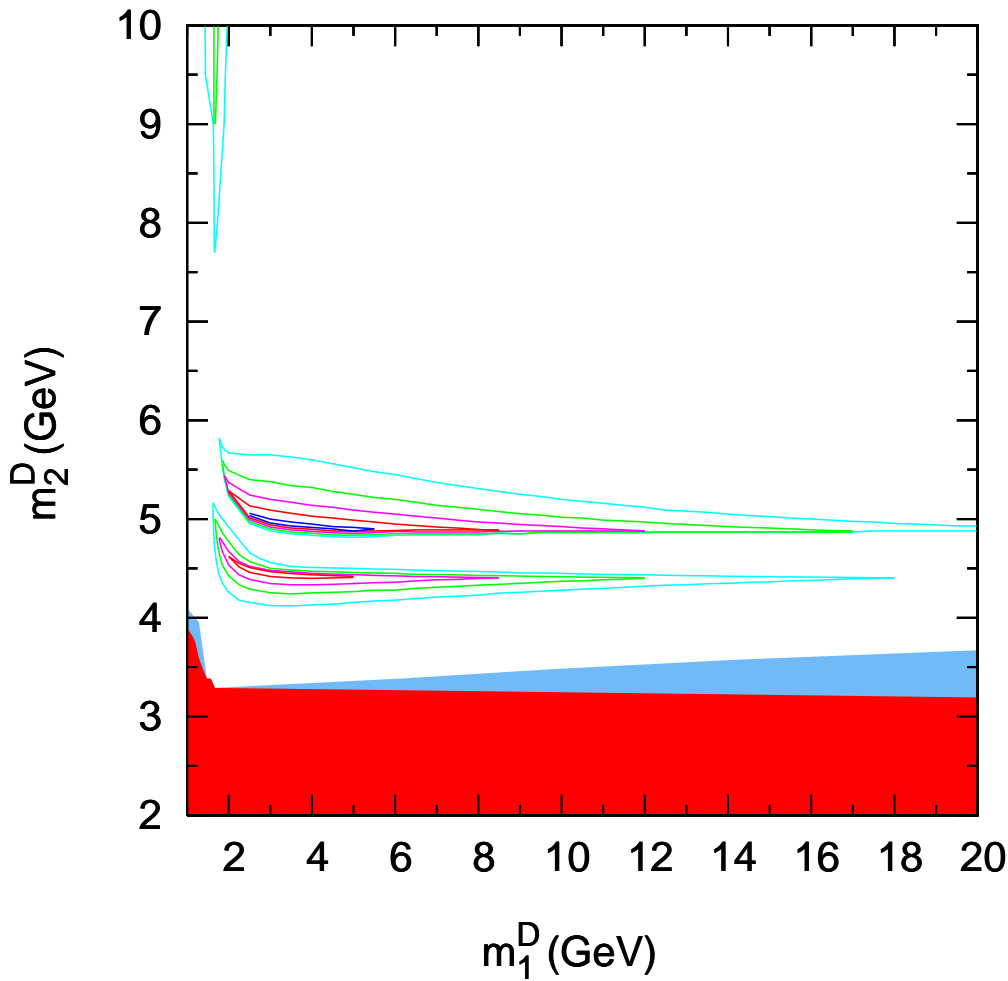}
\caption{The contours with fixed maximal $M_S$
in the $m^D_1-m^D_2$ plane for
$m^D_3=30~{\rm GeV}$. The cyan, green, magenta,
red and blue contours correspond, respectively,
to $M_S=6$, $7$, $8$, $9$ and $10$ in units of
$10^{16}~{\rm GeV}$. In black and white,
contours with higher $M_S$ appear to be
darker. The red/dark shaded area is excluded
by both the requirements that, for the maximal
$M_S$, $M_1>m_{\rm inf}/2$ and $L_3$ is
conserved by dimension-five processes, while
the blue/lightly shaded one only by the
latter.}
\label{fig:MS}
\end{figure*}

\begin{figure*}[t]
\centering
\includegraphics[width=135mm,angle=0]
{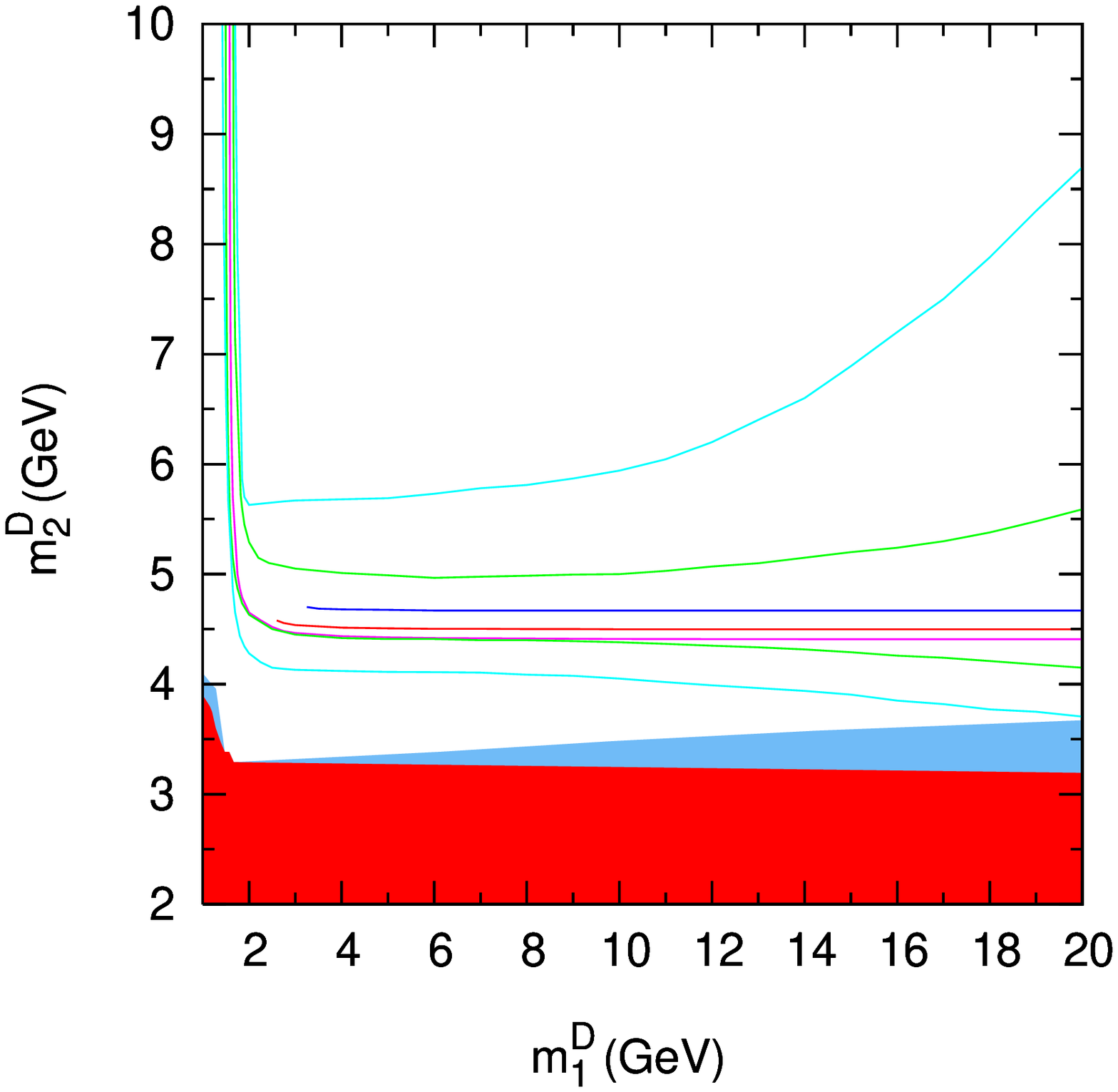}
\caption{The contours with fixed $m_1$
corresponding to the maximal $M_S$ in the
$m^D_1-m^D_2$ plane for $m^D_3=30~{\rm GeV}$.
The cyan, green, magenta, red and blue
contours correspond, respectively, to $m_1=3$,
$5$, $10$, $50$ and $100$ in units of
$10^{-4}~{\rm eV}$. In black and white,
contours with higher $m_1$ appear to be darker.
The red/dark shaded and blue/lightly shaded
areas are as in Fig.~\ref{fig:MS}.}
\label{fig:m1}
\end{figure*}

\par
For definiteness, we set $\tan\beta=50$,
although in this case it scarcely affects the
calculation of baryon asymmetry. We are still
free to vary $M_S$, $m_1$ and the three
neutrino Dirac masses $m^D_i$. As previously
mentioned, we require that all three RHN masses
be greater than $m_{\rm inf}/2$ and less than
$M^2/M_S$, {\it i.e.} $M_1>m_{\rm inf}/2$ and
$M_3M_S<M^2$. By placing one RHN mass $M_j$
near the inflaton pole $M_j=m_{\rm inf}$,
the contribution of the corresponding
self-energy diagram to the baryon asymmetry is
resonantly enhanced (see Eq.~(\ref{Fself})).
However, we regard this as an unnatural
fine-tuning since it apparently cannot be
ensured by any symmetry, and reject solutions
where any $M_j$ is less than 10\% from the
pole, {\it i.e.}
$|M_j-m_{\rm inf}|/m_{\rm inf}<0.1$.
We also impose the requirement that $L_3$ is
not violated by any dimension-five processes
and that the observed baryon asymmetry $n_B/s
\simeq 8.66\times 10^{-11}$, derived from the
recent WMAP data \cite{wmap}, is reproduced.
Under all these restrictions and for fixed
values of $m^D_i$, we find the value of $m_1$
which maximizes $M_S$ (and thus minimizes
$|\hat{\zeta}_1|$ and $|\hat{\zeta}_2|$). We
have chosen to maximize $M_S$ in order to
obtain values as close as possible to the
standard scale
($\approx 5\times 10^{17}~{\rm GeV}$) of the
weakly-coupled heterotic string. We keep only
values of maximal $M_S$ which exceed
$10M\simeq 5.63\times 10^{16}~{\rm GeV}$ so
that the perturbativity requirement is
satisfied.

\par
In Figs.~\ref{fig:MS} and \ref{fig:m1}, we
present the contours with fixed value of this
maximal $M_S$ and the corresponding $m_1$
respectively in the $m^D_1-m^D_2$ plane for
$m^D_3=30~{\rm GeV}$. In Fig.~\ref{fig:MS},
we include only contours with fixed maximal
$M_S$ which satisfies the perturbativity limit.
However, this limit is not necessarily
satisfied everywhere on the contours with fixed
$m_1$ which we show in Fig.~\ref{fig:m1}. It
holds only on the parts of these contours which
lie in the area covered by the contours in
Fig.~\ref{fig:MS}. We observe that
the contours with fixed maximal $M_S$ show the
existence of two main almost horizontal
mountain ranges not far from each other and a
lower almost vertical mountain range at low
values of $m^D_1$. The contours with fixed
$m_1$, on the other hand, form a single
horizontal mountain range and a secondary
almost vertical mountain range at higher values
of $m^D_2$. At low values of $m^D_2$ (and
$m^D_1$), we find that $M_1<m_{\rm inf}/2$ for
a wide range of choices of $m_1$ and our
scenario is irrelevant since the inflaton
decays directly into RHNs. Also, the ratios
$\hat{\zeta}_j/\hat{y}_j$ ($j=1,2$) acquire
large values, which result in the $L_3$
asymmetry being washed out by dimension-five
processes involving a $\hat{\zeta}$ coupling.
The red/dark shaded areas in Figs.~\ref{fig:MS}
and \ref{fig:m1} are excluded because, for the
maximal value of $M_S$, $M_1< m_{\rm inf}/2$
and $L_3$ is violated; the blue/lightly shaded
areas, on the other hand, are excluded only
because of $L_3$ violation. The value of the
maximal $M_S$ in these areas is much smaller
than the perturbativity bound, thus these areas
are not considered acceptable in any case. We
have also constructed the contours in the
$m^D_1-m^D_2$ plane with fixed $M_S$ or $m_1$
for $m^D_3=10~{\rm GeV}$ and
$m^D_3=20~{\rm GeV}$. We find that they reveal
a very similar structure.

\par
\begin{figure}[t]
\centering
\includegraphics[width=62mm,angle=-90]
{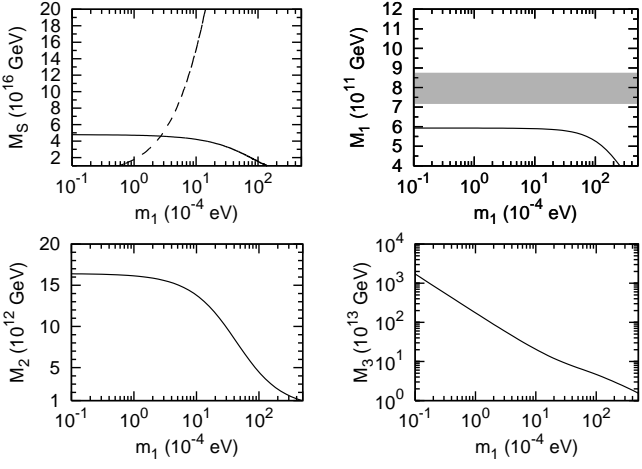}
\caption{The value of $M_S$ required to obtain
the correct baryon asymmetry as a function of
$m_1$ (solid line in the top left panel) for
$m^D_1=7~{\rm GeV}$, $m^D_2=4~{\rm GeV}$ and
$m^D_3=30~{\rm GeV}$. The constraint
$M_3M_S<M^2$ holds to the right of the dashed
line in this panel. The other three panels show
the masses $M_i$ ($i=1,2,3$) as functions of
$m_1$ for the same values of the Dirac neutrino
masses. Regions excluded by the requirement
that $|M_1-m_{\rm inf}|/m_{\rm inf}>0.1$ are
shaded.}
\label{fig:mD2=4}
\end{figure}
\begin{figure}[t]
\centering
\includegraphics[width=62mm,angle=-90]
{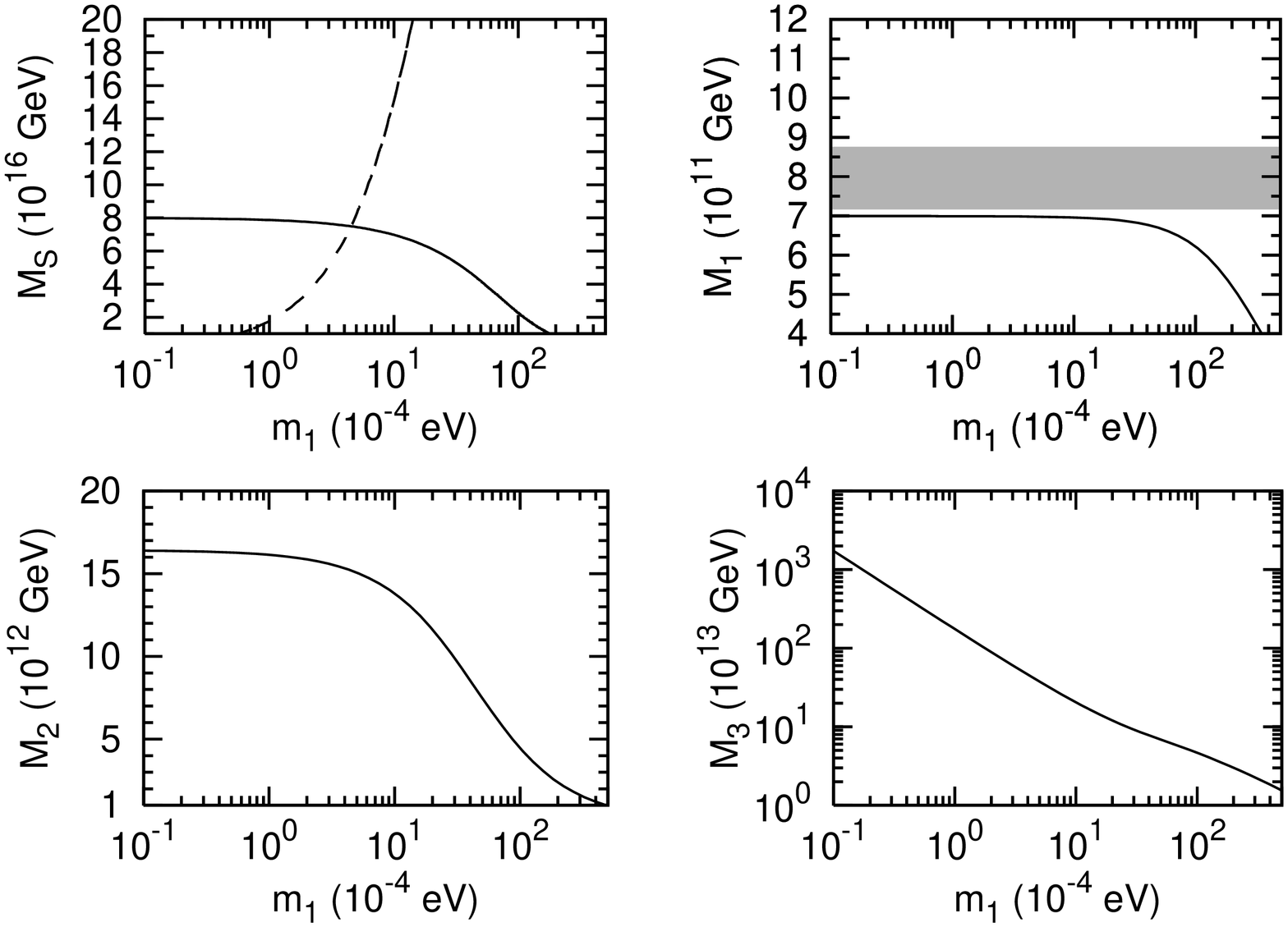}
\caption{As in Fig.~\ref{fig:mD2=4}, but for
$m^D_2=4.35~{\rm GeV}$.}
\label{fig:mD2=4.35}
\end{figure}
\begin{figure}[t]
\centering
\includegraphics[width=62mm,angle=-90]
{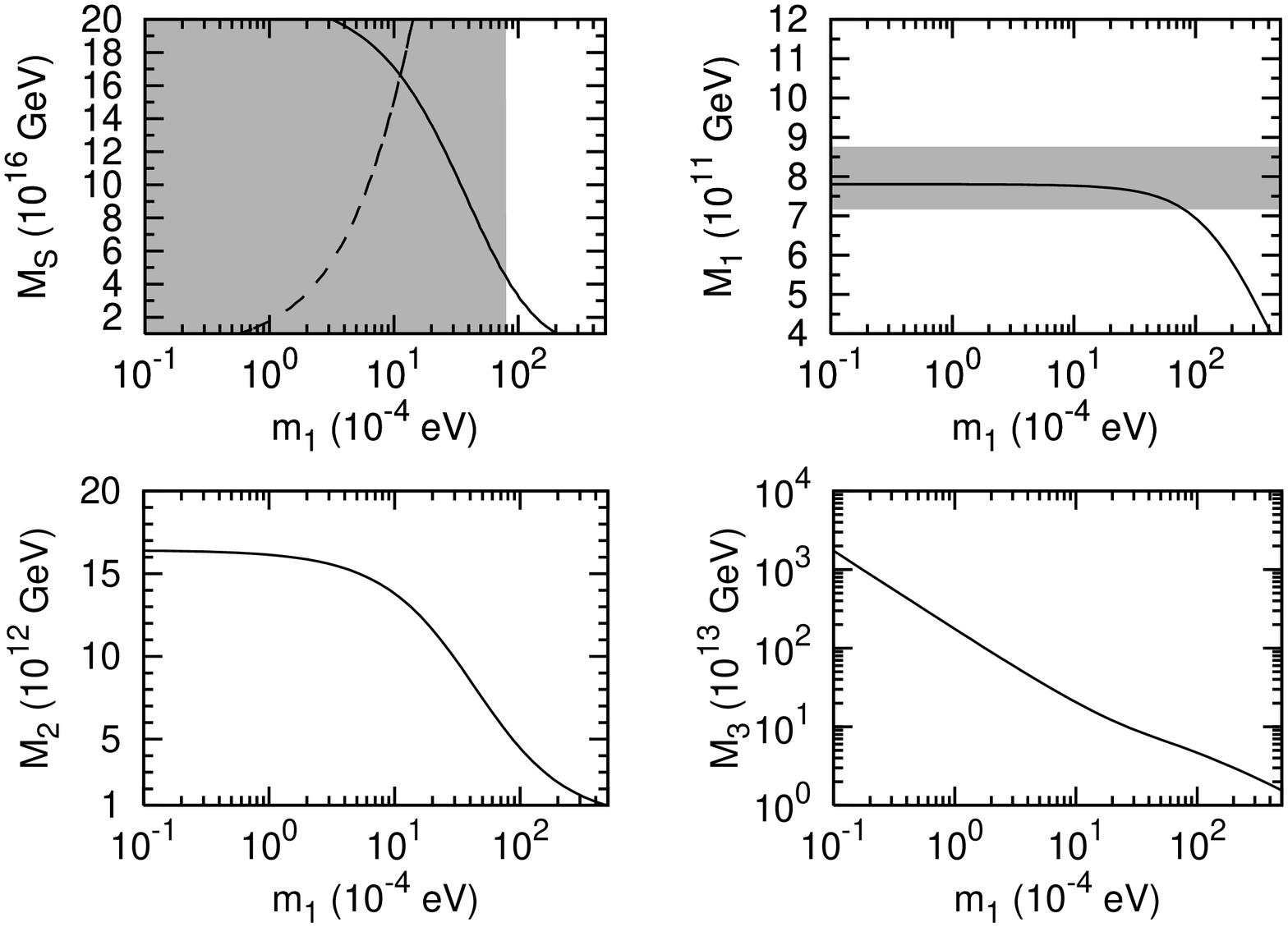}
\caption{As in Fig.~\ref{fig:mD2=4}, but for
$m^D_2=4.6~{\rm GeV}$.}
\label{fig:mD2=4.6}
\end{figure}
\begin{figure}[t]
\centering
\includegraphics[width=62mm,angle=-90]
{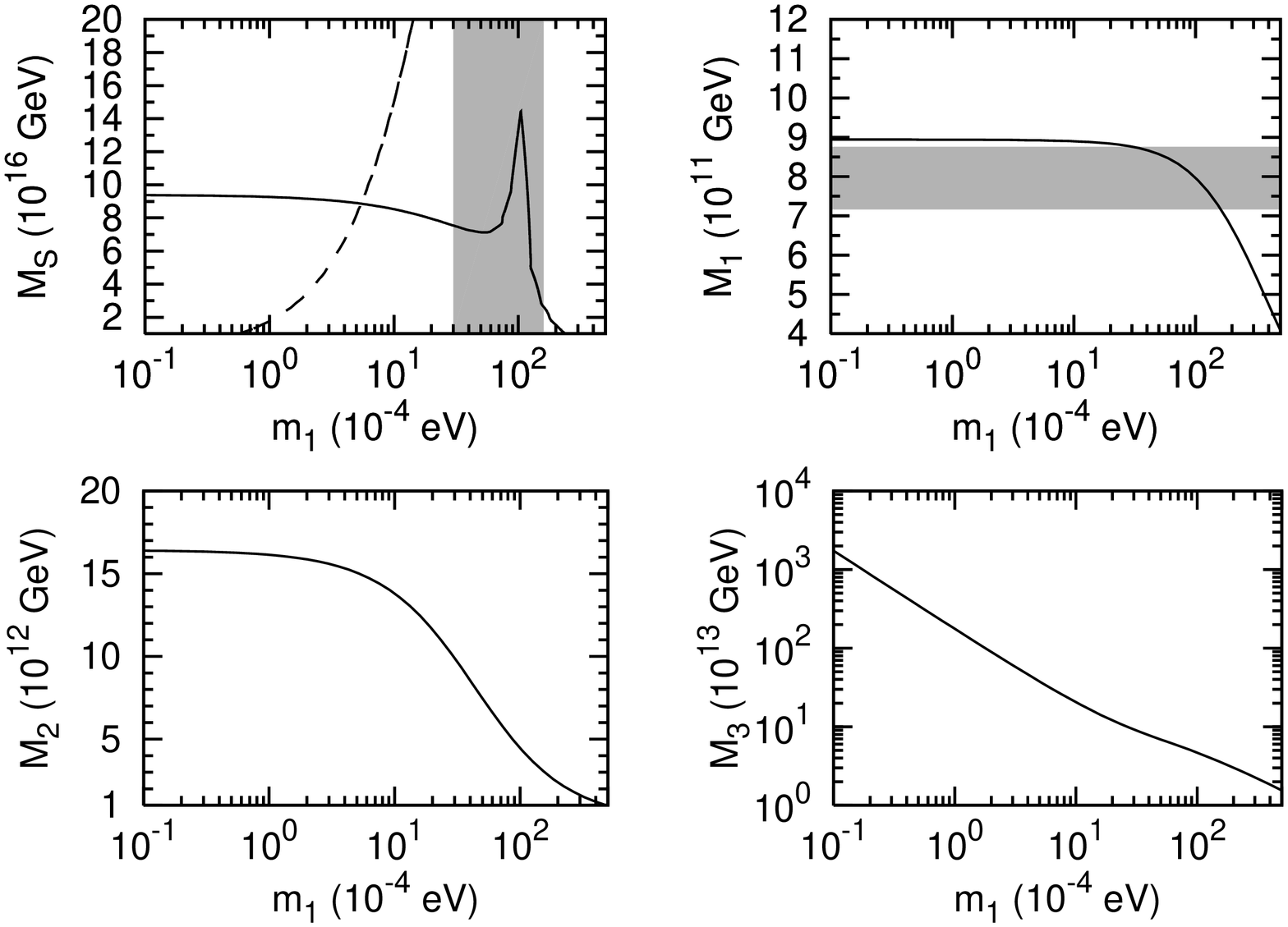}
\caption{As in Fig.~\ref{fig:mD2=4}, but for
$m^D_2=4.95~{\rm GeV}$.}
\label{fig:mD2=4.95}
\end{figure}
\begin{figure}[t]
\centering
\includegraphics[width=62mm,angle=-90]
{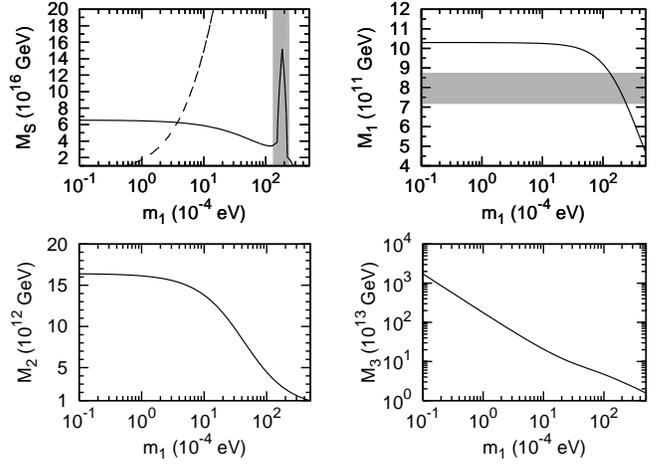}
\caption{As in Fig.~\ref{fig:mD2=4}, but for
$m^D_2=5.3~{\rm GeV}$.}
\label{fig:mD2=5.3}
\end{figure}

In order to understand the structure of these
contour plots, let us consider some specific
values for the three Dirac neutrino masses
$m^D_i$. For instance, let us fix
$m^D_1=7~{\rm GeV}$, $m^D_3=30~{\rm GeV}$ and
take five characteristic choices of $m^D_2$
($4,\,4.35,\,4.6,\,4.95,\,5.3~{\rm GeV}$). Then
compute the value of $M_S$ which is required in
each case to obtain the correct $n_B/s$ as a
function of $m_1$. The result is depicted in
the top left panels of
Figs.~\ref{fig:mD2=4}-\ref{fig:mD2=5.3} by a
solid line. Moreover, we find that the
constraint $M_3M_S<M^2$ is satisfied to the
right of the dashed line in these panels. In
the other three panels in each of these
figures, we show the corresponding values of
$M_i$, $i=1,2,3$, as functions of $m_1$. The
shaded bands are excluded by the requirement
that $|M_1-m_{\rm inf}|/m_{\rm inf}>0.1$.

\par
To appreciate the meaning of these figures, we
must first observe that, as our numerical
findings show, the main contribution to the
baryon asymmetry comes from the interference of
vertex with self-energy diagrams, while the
interference between self-energy diagrams is
subdominant. The heaviest RHN mass $M_3$ is
much larger than $M_1$, $M_2$ and $m_{\rm inf}$
as one can see from
Figs.~\ref{fig:mD2=4}-\ref{fig:mD2=5.3}. So,
the contribution to the baryon asymmetry from
diagrams with $\nu^c_3$ exchange is
suppressed. Moreover, in the interesting range
of parameters where the maximal $M_S$ is
achieved, $M_1$ is nearer the inflaton pole
than $M_2$ is. As a consequence, the dominant
self-energy diagram is the one with $\nu^c_1$
exchange. It also turns out from our numerical
study that the dominant vertex diagram in the
interesting range of parameters is the one with
$\nu^c_2$ exchange. So, the baryon asymmetry is
dominated by the interference of the
self-energy diagram with $\nu^c_1$ exchange
with the vertex diagram with $\nu^c_2$
exchange.

\par
From Figs.~\ref{fig:mD2=4}-\ref{fig:mD2=5.3},
we see that, for $m_1\rightarrow 0$, the masses
$M_1$ and $M_2$ acquire constant asymptotic
values, while $M_3$ keeps increasing. In
consequence, the kinematic factor in the
dominant contribution to $n_B/s$, which
originates from the ``stripped'' diagrams, is
constant in this limit. The prefactor
originating from the product of coupling
constants contains $U^c_{31}$ and $U^c_{32}$
which connect the $\hat{\nu}^c_3$ state with
the $\nu^c_{1,2}$ mass eigenstates. We find
that, as $m_1$ decreases, these elements of
$U^c$ first rise sharply and then approach
their constant asymptotic values. Their rapid
increase is compensated by an appropriate
increase of $M_S$, which enters the prefactor
through $|\hat{\zeta}_1|$ and
$|\hat{\zeta}_2|$, so that the baryon asymmetry
remains equal to its WMAP value. This explains
the fact that, as $m_1$ decreases, $M_S$
generally exhibits a fast increase before
entering its asymptotic plateau (see solid
line in the top left panels of
Figs.~\ref{fig:mD2=4}-\ref{fig:mD2=5.3}). This
phenomenon generally helps us to obtain larger
maximal values of $M_S$.

\par
From these figures, we also see that $M_2$ and
$M_3$ (see two bottom panels) remain
effectively unaltered as we vary $m^D_2$ for
fixed $m^D_1$ and $m^D_3$. This explains also
the fact that the constraint $M_3M_S<M^2$, as
depicted in the top left panels (dashed line),
is practically unaffected by changes in the
value of $m^D_2$. The $M_1$ though is enhanced
as $m^D_2$ increases. For $m^D_2=4~{\rm GeV}$,
we see from Fig.~\ref{fig:mD2=4} that the
asymptotic value of $M_1$
(as $m_1\rightarrow 0$) lies well below the
inflaton pole. So, there is no enhancement of
the dominant ``stripped'' self-energy diagram
produced by this pole. As a
consequence, the values of $M_S$ yielding the
correct baryon asymmetry are relatively low.
So, the maximal $M_S$, which is achieved at
$M_3M_S=M^2$ in this case, is also quite low
and corresponds to a low value of $m_1$
($\sim 10^{-4}~{\rm eV}$). This explains the
fact that this particular case lies well below
the lower horizontal mountain range in
Fig.~\ref{fig:MS} and the horizontal mountain
range in Fig.~\ref{fig:m1}.

\par
Let us now discuss how the situation changes as
we increase $m^D_2$ with $m^D_1$ and $m^D_3$
fixed. For $m^D_2=4.35~{\rm GeV}$,
Fig.~\ref{fig:mD2=4.35} shows that the
asymptotic $M_1$, which is larger than in the
previous case, gets closer to the pole and thus
the dominant ``stripped'' self-energy diagram
is enhanced, leading to a larger
maximal $M_S$. This is again achieved at
$M_3M_S=M^2$ and corresponds to a somewhat
bigger, but still quite small value of $m_1$.
This case lies near the brow of the lower
mountain range in Fig.~\ref{fig:MS}, but still
below the mountain range in Fig.~\ref{fig:m1}.
Increasing $m^D_2$ further to the value
$4.6~{\rm GeV}$, we see from
Fig.~\ref{fig:mD2=4.6} that the asymptotic
$M_1$ gets very close to the pole and, thus,
$M_S$ is further enhanced. However, $M_1$ now
enters into the band excluded by the
requirement
$|M_1-m_{\rm inf}|/m_{\rm inf}>0.1$, which in
the top left panel extends to the right of the
line $M_3M_S=M^2$. So, in this case, the
maximal $M_S$ is achieved at the boundary of
this excluded band and not at $M_3M_S=M^2$. It
is thus quite low and corresponds to a
considerably larger $m_1$. This case lies in
the valley extending between the two horizontal
mountain ranges in Fig.~\ref{fig:MS} and very
near the brow of the horizontal mountain range
in Fig.~\ref{fig:m1}. In
Fig.~\ref{fig:mD2=4.95}, we present the case
$m^D_2=4.95~{\rm GeV}$. We see that the
asymptotic $M_1$ now lies above the excluded
band, but still close to the pole. The
whole excluded band in the top left panel now
lies to the right of the line $M_3M_S=M^2$, and
thus the maximal $M_S$ is again achieved on
this line and corresponds to a low $m_1$.
Therefore, this case lies above the brow of the
horizontal mountain range in Fig.~\ref{fig:m1}.
Actually, at some value of $m^D_2$, the point
at which the maximal $M_S$ is achieved jumps
from the right boundary of the excluded band to
the line $M_3M_S=M^2$. So, $m_1$ drops suddenly
to low values. This discontinuity explains the
absence of the upper branches of the contours
with $m_1=10^{-2}$, $5\times 10^{-3}$ and
$10^{-3}~{\rm eV}$. The enhancement of the
self-energy diagram due to the pole yields the
sharp peak of $M_S$ within the excluded
(shaded) band. The tail of this enhancement
leads to a large value for the maximal $M_S$.
So, this case is near the brow of the upper
mountain range in Fig.~\ref{fig:MS}. For
$m^D_2=5.3~{\rm GeV}$, the asymptotic $M_1$
moves away from the pole and the excluded
(shaded) band in the top left panel further to
the right (see Fig.~\ref{fig:mD2=5.3}).
Consequently, the maximal $M_S$ is still
achieved on the line $M_3M_S=M^2$, but drops to
lower values. This case is well above the upper
horizontal mountain range in Fig.~\ref{fig:MS}
and the horizontal mountain range in
Fig.~\ref{fig:m1}.

\par
The secondary vertical mountain range in
Fig.~\ref{fig:MS} can be understood as follows.
As $m^D_1$ decreases for fixed and large values
of $m^D_2$, the mass $M_1$ also decreases. As a
consequence, the band excluded by the
requirement that
$|M_1-m_{\rm inf}|/m_{\rm inf}>0.1$ moves to
the left in the $m_1-M_S$ plane approaching the
curve $M_3M_S=M^2$ and the maximal $M_S$, which
is achieved on this curve, increases since the
inflaton pole gets nearer. At some small value
of $m^D_1$, the left boundary of this band
touches the point corresponding to the maximal
$M_S$, which thus reaches its largest possible
value for the chosen value of $m^D_2$. This
yields the brow of the vertical mountain range
in Fig.~\ref{fig:MS}. The corresponding value
of $m_1$ is, however, still relatively small.
For smaller values of $m^D_1$, the band moves
further to the left and the point of maximal
$M_S$ jumps suddenly to the right boundary of
the band. Thus, the maximal $M_S$ drops, while
the corresponding $m_1$ rises sharply forming
the vertical mountain range in
Fig.~\ref{fig:m1}. This sharp discontinuity in
$m_1$ explains the absence of the right branch
of the contour with $m_1=10^{-3}~{\rm eV}$. As
the value of $m^D_1$ decreases further, the
excluded band shifts further to the left and
the maximal $M_S$ and the corresponding $m_1$
drop smoothly.

\par
Despite the abundance of adjustable parameters
in the non-${\rm SU}(2)_R$-symmetric case, it
turns out not to be easy to achieve the
observed baryon asymmetry. In particular, we
find that, for central values of the neutrino
mass-squared differences and mixing angles, the
string scale $M_S$ is restricted to be lower
than about $10^{17}~{\rm GeV}$. So, the
standard weakly-coupled heterotic string scale
($\approx 5\times 10^{17}~{\rm GeV}$) cannot be
achieved in this case. However, lower string
scales, such as the ones encountered here, can
be easily obtained in other string models like
the strongly-coupled heterotic string from
M-theory (see {\it e.g.}
Ref.~\cite{stringscale}). Also, we find that it
is difficult to generate an adequate baryon
asymmetry with
$T_{\rm reh}\sim 10^9~{\rm GeV}$, which is the
standard upper bound on $T_{\rm reh}$ from the
gravitino constraint \cite{khlopov,gravitino}.
We are thus obliged to allow higher reheat
temperatures, which is \cite{gravitino} though
perfectly possible provided that the gravitino
decay to photons and photinos is somewhat
suppressed.

\par
At the upper bound on $M_S$ ($\approx 10^{17}~
{\rm GeV}$), the parameters $|\hat{\zeta}_i|$
($i=1,2$) are about $5.63\times 10^{-2}$ and
may thus lead to processes observable in the
future colliders. (For smaller values of $M_S$,
they are even larger.) Indeed, as mentioned in
Sec.~\ref{sec:rsym}, the explicit R-parity
violation in our model, required for
leptogenesis, has some low-energy signatures
coming from dimension-four effective scalar
vertices. They may typically be the three-body
slepton decay processes
\beq
\tilde{l}_i\rightarrow h_1h_2h_2^*,
\label{decay}
\eeq
for $i=1,2$, which can easily be kinematically
allowed. The magnitude of their effective
coupling constants $\hat{\zeta}_i\hat{y}^*_i$
is of order $10^{-3}$ near the largest possible
value of $M_S$, which is achieved at
$m^D_1\approx 3~{\rm GeV}$ and
$m^D_2\approx 5~{\rm GeV}$ for
$m^D_3=30~{\rm GeV}$ (see Fig.~\ref{fig:MS}).
The corresponding decay rates are then of order
$10^{-8}~{\rm GeV}$ for mass of the decaying
slepton of order $1~{\rm TeV}$.

\par
We have seen that the maximal value of $M_S$,
in most cases, is achieved at $M_3M_S=M^2$, and
the corresponding value of $m_1$ lies in the
range $10^{-4}-10^{-3}~{\rm eV}$. However,
for values of the Dirac neutrino masses in the
valley between the two horizontal mountain
ranges appearing in the $M_S$ contour plots in
the $m^D_1-m^D_2$ plane for fixed $m^D_3$ (see
Fig.~\ref{fig:MS}), the maximal $M_S$ is
achieved at the boundary of the band excluded
by the requirement
$|M_1-m_{\rm inf}|/m_{\rm inf}>0.1$ and the
corresponding $m_1$ is considerably larger
reaching values of order $10^{-2}~{\rm eV}$.
This case, however, yields very low maximal
values of $M_S$ and must, therefore, be
excluded by the perturbativity requirement.
In conclusion, we predict that, for acceptable
values of $M_S$, the (in principle observable)
smallest neutrino mass $m_1$ takes values
between $10^{-4}$ and $10^{-3}~{\rm eV}$. Thus
our assumption of a hierarchical light neutrino
spectrum is self-consistent.

\subsection{${\rm SU}(2)_R$-symmetric case}

\par
In the ${\rm SU}(2)_R$-symmetric case, we again
put $\kappa=10^{-4}$. The COBE value of the
quadrupole anisotropy of the CMBR is now
reproduced for $\lambda\simeq 3.45\times
10^{-4}$ and $M\simeq 5.69\times 10^{15}~
{\rm GeV}$, yielding $m_{\rm inf}\simeq 8.05
\times 10^{11}~{\rm GeV}$. The spectral index
is again equal to unity for all practical
purposes. In this case, $\tan\beta\simeq 55$
and the asymptotic values of $\hat{y}_i$ are
$0.00014$, $0.028$ and $0.66$ for $i=1,2$ and 3
respectively, as already mentioned. So, in the
${\rm SU}(2)_R$-symmetric case, only $M_S$ and
$m_1$ remain free.

\par
To give a chance to this very restrictive case
to possibly yield acceptable values of the
baryon asymmetry, we allow all the neutrino
oscillation parameters to vary within their
$2\sigma$ confidence interval given in
Ref.~\cite{maltoni} (for simplicity, we take
the Dirac phase $\delta=0$). Even
then, we find that, when the restriction
$\hat{\zeta}_3=0$ from preservation of the
$B-3L_3$ asymmetry is imposed, the resulting
values of the baryon asymmetry are always some
orders of magnitude smaller than the observed
value for any reasonable values of $M_S$
(satisfying the perturbativity requirement) and
any $m_1$ in the allowed range from WMAP data
\cite{wmap}.

\par
The main reason for this failure is the
smallness of the off-diagonal elements of
$U^c$. The existence of a CP asymmetry requires
that $\hat{y}_3$ vertices be connected to
$\hat{\zeta}_1$ or $\hat{\zeta}_2$ vertices
through $\nu^c$ internal lines, yielding
products of the type $U^c_{3j}U^{c\ast}_{3k}
U^c_{1j}U^{c\ast}_{2k}$ for fixed values of $j$
and $k$ ($j\neq k$). These products contain at
least two off-diagonal elements of $U^c$.
However, the strongly hierarchical values of
$\hat{y}_i$ resulting from the ${\rm SU}(2)_R$
symmetry imply that the off-diagonal elements
of $U^c$ are always small. In addition, the
requirement from the preservation of the
$B-3L_3$ asymmetry that
$|U^c_{31}|\lesssim 10^{-5}$ further restricts
this product of matrix entries, such that the
overall product of coupling constants entering
the $L_3$ asymmetry per inflaton decay
($\epsilon_3$) is already less than $10^{-8}$,
without considering any loop suppression
factors. In contrast, in the
non-${\rm SU}(2)_R$-symmetric case, the values
of the off-diagonal elements of $U^c$ can be of
order $0.1$ or even unity.

\section{Conclusions}
\label{sec:concl}

\par
We proposed a scenario of non-thermal
leptogenesis following SUSY hybrid inflation,
in the case where the light neutrinos acquire
masses exclusively via the standard seesaw
mechanism, {\it i.e.} through their coupling to
heavy RHNs, and the decay of the inflaton to
RHN superfields is kinematically blocked
or (in the presence of a ${\rm SU}(2)_R$ gauge
symmetry) the RHNs with mass smaller than half
the inflaton mass are too light to generate
sufficient baryon asymmetry through their
subsequent decay. The primordial lepton
asymmetry is generated through the direct decay
of the inflaton into light particles. We
explored our scenario within the context of two
simple SUSY GUT models, one without and one with
${\rm SU}(2)_R$ gauge symmetry, which
incorporate the standard version of SUSY hybrid
inflation.

\par
The $\mu$ problem is solved via a ${\rm U}(1)$
R-symmetry which forbids the existence of an
explicit $\mu$ term, while allows a trilinear
superpotential coupling of the gauge singlet
inflaton superfield to the electroweak Higgs
superfields. After the spontaneous breaking of
the GUT gauge symmetry, this singlet inflaton
acquires a suppressed VEV due to the soft
SUSY-breaking terms. Its trilinear coupling to
the Higgs superfields then yields a $\mu$ term
of the right magnitude.

\par
The main decay mode of the inflaton is to a
pair of electroweak Higgs superfields via the
same trilinear coupling. The initial lepton
asymmetry is created in the subdominant decay
of the inflaton to a lepton and an electroweak
Higgs superfield via the interference of
one-loop diagrams with exchange of different
RHNs. The existence of these diagrams requires
the presence of some specific superpotential
couplings which explicitly violate the
${\rm U}(1)$ R-symmetry and R-parity. These
couplings, however, do not affect the exact
baryon number conservation in perturbation
theory which is implied by the R-symmetry.
Thus, the only way to generate baryons in
these models is via a primordial leptogenesis
(or via electroweak baryogenesis).

\par
In our analysis, we took into account the
constraints from neutrino masses and mixing.
The requirement that the primordial lepton
asymmetry not be erased by
lepton-number-violating processes before the
electroweak phase transition is a much more
stringent constraint on the parameters of the
theory. We showed that the model with
${\rm SU}(2)_R$ gauge symmetry is too
restrictive to be able to generate an adequate
baryon asymmetry in accordance with these
constraints and for natural values of the other
parameters even if we allow the neutrino
oscillation parameters to vary within their
$2\sigma$ confidence intervals. It is thus
ruled out.

\par
On the contrary, the
non-${\rm SU}(2)_R$-symmetric model can be
viable even with central values of the neutrino
mass-squared differences and mixing angles.
However, we find that this model is much more
restrictive than the model studied in
Ref.~\cite{previous}, which contained
${\rm SU}(2)_L$ triplet superfields giving a
second contribution to light neutrino masses
after that of the RHNs. Indeed, in order to
generate the observed BAU, we had to take a
larger reheat temperature, which is though
perfectly acceptable if the gravitino decay to
photons and photinos is somewhat suppressed,
and a string scale somewhat smaller than the
weakly-coupled heterotic string one. Such lower
string scales are easily obtained in other
string models such as the strongly-coupled
heterotic string from M-theory. We also find
that the lightest neutrino mass eigenvalue,
which is an in principle measurable parameter,
is restricted to lie in the range
$10^{-4}-10^{-3}~{\rm eV}$. So, our model
is consistent with a hierarchical neutrino
mass spectrum. The explicit breaking of
R-parity, which is necessary for our
baryogenesis mechanism, need not have currently
observable low-energy signatures, although it
may have signatures detectable in future
colliders. Also, the LSP can be made long-lived
and, thus, be a possible candidate for the CDM
in the universe.

\par
There are many constraints and necessary
conditions for our mechanism to successfully
produce the observed BAU (in the
non-${\rm SU}(2)_R$-symmetric case). Some of
these constraints are generic and easily
satisfied, but some restrict certain {\em a
priori} adjustable parameters of the model to
lie within narrow ranges. This is not a
technical fine-tuning problem, but
might be regarded as an undesirable feature
since we have not presented an underlying
theory which selects these particular ranges of
values. However, given the large number of
possible fundamental theories, it is plausible
that parameter values consistent with our
baryogenesis mechanism will emerge naturally
from one or many of them. For such underlying
theories, which on the basis of previous
baryogenesis scenarios one might think were
ruled out, we have shown that in fact they
may be consistent with data. Furthermore, the
scenario is especially predictive and is more
readily testable compared to baryogenesis
models in which parameter values are not
particularly restricted.

\par
The conditions for the baryogenesis mechanism
to be effective involve the following
parameters: the dimensionless (complex)
effective coupling constants $\hat{\zeta}_i$,
the lightest neutrino mass, the RHN masses, the
neutrino Yukawa coupling constants (including
their complex phases), the inflaton mass and
the reheat temperature. Recall that, for
simplicity, the light neutrino mass-squared
differences and mixing angles are taken to
coincide with their best-fit values and the
complex phases in the MNS matrix are set to
zero. In addition, the model is restricted to
be consistent with cosmological observations
determining the values of the inflationary
parameters and restricting the reheat
temperature and the mass and couplings of the
dark matter candidate.

\par
The first necessary condition is that all RHN
masses should exceed half the inflaton mass.
This is not a fine-tuning. Indeed, it may be
generic in many classes of models. The
hierarchical light neutrino mass spectrum is
similarly restricted to be of the normal
(non-inverted) hierarchical type. We imposed a
condition that RHN masses should not be too
close (within $10\%$) to the inflaton mass
(Sec.~\ref{sec:numernonSU2R}) specifically in
order to avoid regimes where the baryon
asymmetry is due to a resonant enhancement.
These regimes would in principle be allowed,
but the resonance condition appears
coincidental since the RHN masses and the
inflaton mass arise from different sources.
We could remove this condition and the allowed
regions would be slightly larger, but the
explanation of the observed baryon asymmetry
would be more complicated in the resonant
regions, hence for simplicity we do not
consider them.

\par
The ${\rm U}(1)_R$-breaking coupling constants
$\hat{\zeta}_i$ are fixed within relatively
narrow ranges of values. One of them,
$\hat{\zeta}_3$, must be quite small (of order
$10^{-7}$) or vanish in order not to wash out
the produced lepton asymmetry in the $L_3$
direction (Sec.~\ref{sec:preserve}). This
might, for example, be ensured by imposing a
symmetry. The other two must take values of a
few times $10^{-2}$. The upper bound on them
arises from the perturbativity limit on
$M/M_S$, while the lower from the requirement
to produce sufficient $n_B/s$. We have taken
these two coupling constants equal in magnitude
for simplicity, but they may be unequal without
much affecting the success of the model. Since
these values are not particularly tiny and are
not required to cancel against any other
quantity, the restrictions represent an in
principle testable prediction rather than a
fine-tuning. The relative complex phase between
$\hat{\zeta}_1$ and $\hat{\zeta}_2$ is of order
unity.

\par
The neutrino Yukawa coupling constant matrix
(or Dirac mass matrix after ${\rm SU}(2)_L$
breaking) is taken to have small or vanishing
off-diagonal elements, when written in the
``hatted'' weak interaction basis where lepton
family numbers are defined (the charged lepton
mass basis). Three off-diagonal elements can be
set to zero by redefining the RHN superfields,
two of the remaining three must be small or
zero to prevent washout of the $L_3$ asymmetry
(Sec.~\ref{sec:preserve}), and the remaining
one ($\hat{y}_{\nu 12}$) is set to zero for
simplicity. These requirements are not
fine-tuned since the off-diagonal
$\hat{y}_{\nu ij}$'s may be small as a result
of a symmetry, as in the quark sector.

\par
Cosmological parameters that are fixed by
cosmological considerations and data are the
number of e-foldings of our present horizon
scale and the quadrupole anisotropy of the
CMBR, which together determine $M$, and one of
$\kappa$ or $\lambda$ (Sec.~\ref{sec:model}).
The reheat temperature is also constrained
(Sec.~\ref{sec:numerics}) not to overproduce
LSPs from gravitino decay, and (less
stringently) by the usual gravitino constraint
on dissociation of light elements. Since the
lepton asymmetry is proportional to the reheat
temperature, we choose the reheat temperature
at or near its maximum value, which fixes the
remaining one of $\kappa$ and $\lambda$. As a
consequence, the inflaton mass, which enters
into the baryon asymmetry formula, is also
fixed. These values represent neither
predictions nor fine-tunings, but simply
observational constraints on the model. The
conditions that the LSP be long-lived and the
gravitino decays not disrupt primordial
nucleosynthesis both restrict the spectrum of
superpartners, but do not directly influence
the baryogenesis scenario (except through the
effect of the LSP mass on the maximal reheat
temperature).

\par
The remaining parameters are $m^D_i$, the
diagonal Dirac neutrino masses, corresponding
to the neutrino Yukawa coupling constants.
Although the value of $m^D_3$ can be varied
over a range of a few tens of GeV without much
change, the model is rather sensitive to
$m^D_{1,2}$, as can be seen in
Figs.~\ref{fig:MS} and \ref{fig:m1}. The reason
for this is, roughly, that the correct
primordial lepton asymmetry can be obtained
only when one RHN mass is not very far from
$m_{\rm inf}$. This approximate condition
results in a surface in the space of $m^D_i$
near which the model is successful (see
Sec.~\ref{sec:numerics} for a full discussion).
Since $m^D_{1,2}$ are protected by chiral
symmetry, their values are not subject to a
technical fine-tuning problem, and their rather
narrow allowed ranges can be said to be a
prediction of the model, which is in principle
testable. It might be argued that it is
undesirable for the baryon asymmetry to vary
steeply when the basic parameters of the model
are changed by a small amount; and in fact we
have excluded the ``resonance'' regions where
one RHN mass is too close to $m_{\rm inf}$,
where this steep variation does occur. Note,
though, that we do not predict $n_B/s$, but
rather use its known value as a constraint on
the model.

\section*{ACKNOWLEDGEMENTS}
\par
We thank B.C. Allanach for helping us with his
code {\tt SOFTSUSY} \cite{SoftSusy} and T. Hahn
for his help with the software packages of
Ref.~\cite{hahn}. This work was supported by
the European Union under the contract
MRTN-CT-2004-503369. The research of R. Ruiz de
Austri was also supported by the program
``Juan de la Cierva'' of the Ministerio de
Educaci\'{o}n y Ciencia of Spain.

\def\ijmp#1#2#3{{Int. Jour. Mod. Phys.}
{\bf #1},~#3~(#2)}
\def\plb#1#2#3{{Phys. Lett. B }{\bf #1},~#3~(#2)}
\def\zpc#1#2#3{{Z. Phys. C }{\bf #1},~#3~(#2)}
\def\prl#1#2#3{{Phys. Rev. Lett.}
{\bf #1},~#3~(#2)}
\def\rmp#1#2#3{{Rev. Mod. Phys.}
{\bf #1},~#3~(#2)}
\def\prep#1#2#3{{Phys. Rep. }{\bf #1},~#3~(#2)}
\def\prd#1#2#3{{Phys. Rev. D }{\bf #1},~#3~(#2)}
\def\npb#1#2#3{{Nucl. Phys. }{\bf B#1},~#3~(#2)}
\def\npps#1#2#3{{Nucl. Phys. B (Proc. Sup.)}
{\bf #1},~#3~(#2)}
\def\mpl#1#2#3{{Mod. Phys. Lett.}
{\bf #1},~#3~(#2)}
\def\arnps#1#2#3{{Annu. Rev. Nucl. Part. Sci.}
{\bf #1},~#3~(#2)}
\def\sjnp#1#2#3{{Sov. J. Nucl. Phys.}
{\bf #1},~#3~(#2)}
\def\jetp#1#2#3{{JETP Lett. }{\bf #1},~#3~(#2)}
\def\app#1#2#3{{Acta Phys. Polon.}
{\bf #1},~#3~(#2)}
\def\rnc#1#2#3{{Riv. Nuovo Cim.}
{\bf #1},~#3~(#2)}
\def\ap#1#2#3{{Ann. Phys. }{\bf #1},~#3~(#2)}
\def\ptp#1#2#3{{Prog. Theor. Phys.}
{\bf #1},~#3~(#2)}
\def\apjl#1#2#3{{Astrophys. J. Lett.}
{\bf #1},~#3~(#2)}
\def\n#1#2#3{{Nature }{\bf #1},~#3~(#2)}
\def\apj#1#2#3{{Astrophys. J.}
{\bf #1},~#3~(#2)}
\def\anj#1#2#3{{Astron. J. }{\bf #1},~#3~(#2)}
\def\apjs#1#2#3{{Astrophys. J. Suppl.}
{\bf #1},~#3~(#2)}
\def\mnras#1#2#3{{MNRAS }{\bf #1},~#3~(#2)}
\def\grg#1#2#3{{Gen. Rel. Grav.}
{\bf #1},~#3~(#2)}
\def\s#1#2#3{{Science }{\bf #1},~#3~(#2)}
\def\baas#1#2#3{{Bull. Am. Astron. Soc.}
{\bf #1},~#3~(#2)}
\def\ibid#1#2#3{{\it ibid. }{\bf #1},~#3~(#2)}
\def\cpc#1#2#3{{Comput. Phys. Commun.}
{\bf #1},~#3~(#2)}
\def\astp#1#2#3{{Astropart. Phys.}
{\bf #1},~#3~(#2)}
\def\epjc#1#2#3{{Eur. Phys. J. C}
{\bf #1},~#3~(#2)}
\def\nima#1#2#3{{Nucl. Instrum. Meth. A}
{\bf #1},~#3~(#2)}
\def\jhep#1#2#3{{J. High Energy Phys.}
{\bf #1},~#3~(#2)}
\def\lnp#1#2#3{{Lect. Notes Phys.}
{\bf #1},~#3~(#2)}
\def\appb#1#2#3{{Acta Phys. Polon. B}
{\bf #1},~#3~(#2)}
\def\njp#1#2#3{{New J. Phys.}
{\bf #1},~#3~(#2)}
\def\pl#1#2#3{{Phys. Lett. }{\bf #1B},~#3~(#2)}


\begin{thebibliography}{99}

\bibitem{lepto}
M. Fukugita and T. Yanagida, Phys. Lett. B
{\bf 174}, 45 (1986).

\bibitem{tripletdecay}
P.J. O'Donnell and U. Sarkar, Phys. Rev. D
{\bf 49}, 2118 (1994); 
E. Ma and U. Sarkar, Phys. Rev. Lett.
{\bf 80}, 5716 (1998).

\bibitem{seesaw}
T. Yanagida, in {\it Proceedings of the
Workshop on Unified Theories and Baryon
Number in the Universe}, edited by A.
Sawada and A. Sugamoto (KEK Rep. No. 79-18,
Tsukuba, Japan, 1979), p. 95;
S.L. Glashow, in {\it Quarks and Leptons,
Carg\'{e}se 1979}, edited by M. L\'{e}vy
{\it et al.} (Plenum, New York, 1980), p.
707; M. Gell-Mann, P. Ramond and R. Slansky,
in {\it Supergravity}, edited by P. Van
Nieuwenhuizen and D.Z. Freedman (North
Holland, Amsterdam, 1979), p. 315;
R.N. Mohapatra and G. Senjanovi\'{c},
\prl{44}{1980}{912}.

\bibitem{triplet}
G. Lazarides, Q. Shafi and C. Wetterich,
\npb{181}{1981}{287};
J. Schechter and J.W.F. Valle,
\prd{22}{1980}{2227};
R.N. Mohapatra and G. Senjanovi\'{c},
\prd{23}{1981}{165};
C. Wetterich, \npb{187}{1981}{343};
J. Schechter and J.W.F. Valle,
\prd{25}{1982}{774}.

\bibitem{khlopov}
M.Yu. Khlopov and A.D. Linde,
Phys. Lett. {\bf 138B}, 265 (1984);
J. Ellis, J.E. Kim and D. Nanopoulos,
\ibid{145B}{1984}{181};
I.V. Falomkin, D.B. Pontecorvo, M.G.
Sapozhnikov, M.Yu. Khlopov, F. Balestra
and G. Piragino, \sjnp{39}{1984}{626}.

\bibitem{gravitino}
J.R. Ellis, D.V. Nanopoulos and S. Sarkar,
\npb{259}{1985}{175};
J.R. Ellis, G.B. Gelmini, J.L. L\'{o}pez,
D.V. Nanopoulos and S. Sarkar,
\ibid{B373}{1992}{399}.

\bibitem{pilaftsis}
A. Pilaftsis, \prd{56}{1997}{5431}.

\bibitem{deg}
T. Hambye, E. Ma and U. Sarkar, Nucl. Phys.
{\bf B602}, 23 (2001);
J.R. Ellis, M. Raidal and T. Yanagida,
\plb{546}{2002}{228};
A. Pilaftsis and T.E.J. Underwood,
\npb{692}{2004}{303}.

\bibitem{inflepto}
G. Lazarides and Q. Shafi,
\plb{258}{1991}{305};
G. Lazarides, C. Panagiotakopoulos and
Q. Shafi, \ibid{315}{1993}{325};
{\bf 317}, 661(E) (1993).

\bibitem{nonthtripletdec}
G. Lazarides, hep-ph/9905450.

\bibitem{previous}
T. Dent, G. Lazarides and R. Ruiz de Austri,
\prd{69}{2004}{075012}.

\bibitem{allahv}
R. Allahverdi and A. Mazumdar, Phys. Rev. D
{\bf 67}, 023509 (2003).

\bibitem{raidal}
J.R. Ellis, M. Raidal and T. Yanagida,
\plb{581}{2004}{9}.

\bibitem{cllsw}
E.J. Copeland, A.R. Liddle, D.H. Lyth,
E.D. Stewart and D. Wands,
\prd{49}{1994}{6410}.

\bibitem{dss}
G.R. Dvali, Q. Shafi and R.K. Schaefer,
\prl{73}{1994}{1886}.

\bibitem{hybrid}
A.D. Linde, \plb{259}{1991}{38};
\prd{49}{1994}{748}.

\bibitem{lr}
G.R. Dvali, G. Lazarides and Q. Shafi,
\plb{424}{1998}{259}.

\bibitem{wmap}
C.L. Bennett {\it et al.},
\apjs{148}{2003}{1};
D.N. Spergel {\it et al.},
\ibid{148}{2003}{175}.

\bibitem{lectures}
G. Lazarides, \lnp{592}{2002}{351};
hep-ph/0204294.

\bibitem{mixing}
Of course, there should be corrections to the
equality $y_{uij}=y_{dij}$ to account for quark
mixing.

\bibitem{atmo}
G. Lazarides and N.D. Vlachos,
Phys. Lett. B {\bf 441}, 46 (1998).

\bibitem{atmotalk}
G. Lazarides, hep-ph/9904372.

\bibitem{largetanb}
G. Lazarides and C. Panagiotakopoulos,
\plb{337}{1994}{90}.

\bibitem{maltoni}
M. Maltoni, T. Schwetz, M.A. T\'{o}rtola and
J.W.F. Valle, \njp{6}{2004}{122}.

\bibitem{notation}
When discussing reactions, diagrams, Lagrangian
terms and chemical potentials, we take the $l$
superfields in the ``hatted'' basis; however,
in these cases, we suppress the hats for
simplicity of notation.

\bibitem{covi}
L. Covi, E. Roulet and F. Vissani,
Phys. Lett. B {\bf 384}, 169 (1996).

\bibitem{Pilaftsisren}
B.A. Kniehl and A. Pilaftsis, Nucl. Phys.
{\bf B474}, 286 (1996);
A. Pilaftsis, \ibid{B504}{1997}{61}.

\bibitem{dilog}
L.G. Cabral-Rosetti and M.A. Sanchis-Lozano,
hep-ph/0206081.

\bibitem{DreinerR}
H.K. Dreiner and G.G. Ross, Nucl. Phys.
{\bf B410}, 188 (1993).

\bibitem{turner}
J.A. Harvey and M.S. Turner,
Phys. Rev. D {\bf 42}, 3344 (1990).

\bibitem{lss}
G. Lazarides, R.K. Schaefer and Q. Shafi,
\prd{56}{1997}{1324}.

\bibitem{Glashow}
For an inverse hierarchical spectrum, this
limit also holds (see {\it e.g.} P.H. Frampton,
S.L. Glashow and D. Marfatia,
\plb{536}{2002}{79}).

\bibitem{IbanezQ}
L.E. Ib{\' a}{\~ n}ez and F. Quevedo,
Phys. Lett. B {\bf 283}, 261 (1992).

\bibitem{error}
Note that the signs for the gaugino
contributions in the corresponding relations of
Ref.~\cite{IbanezQ} are incorrect, which led to
erroneous results.

\bibitem{bcw}
C. Bal\'{a}zs, M. Carena and C.E.M. Wagner,
\prd{70}{2004}{015007}.

\bibitem{goldstone}
This may be confirmed by considering that the
negatively charged would-be Goldstone mode
$h^-$ must have the same chemical potential as
the $W^-$ in order to constitute its
longitudinal mode. From the relation
$h^-= W^--h$, implied by the ${\rm SU}(2)_L$
gauge interaction, we then conclude that $h=0$.

\bibitem{km}
M. Kawasaki and T. Moroi, Prog. Theor. Phys.
{\bf 93}, 879 (1995).

\bibitem{cobe}
C.L. Bennett {\it et al.}, \apj{464}{1996}{L1}.

\bibitem{stringscale}
D.G. Cerde\~{n}o and C. Mu\~{n}oz, Phys. Rev. D
{\bf 61}, 016001 (2000);
\ibid{66}{2002}{115007}.

\bibitem{SoftSusy}
B.C. Allanach, Comput.\, Phys.\, Commun.
{\bf 143}, 305 (2002).

\bibitem{hahn} T. Hahn,
\appb{30}{1999}{3469}.

\end{thebibliography}
\end{document}